\pdfoutput=1
\documentclass[10pt,twocolumn,letterpaper]{article}

\usepackage[pagenumbers]{cvpr}
\usepackage{multirow}
\usepackage{colortbl}
\usepackage{tabularx}
\definecolor{oursrow}{RGB}{229,239,252}

\definecolor{cvprblue}{rgb}{0.21,0.49,0.74}
\usepackage[pagebackref,breaklinks,colorlinks,allcolors=cvprblue]{hyperref}

\makeatletter
\renewcommand\paragraph{\@startsection{paragraph}{4}{\z@}{1.5ex \@plus .5ex \@minus .2ex}{-1em}{\normalfont\normalsize\bfseries}}
\gdef\@belowtextlist{}
\newcommand{\belowtext}{\begingroup\let\@elt\relax\xdef\@belowtextlist{\@belowtextlist\@elt\@currbox}\endgroup}
\def\@checkbelowtext#1{\def\@elt##1{\ifx##1#1\@tempswatrue\fi}\@belowtextlist\let\@elt\@checkbelowtext}
\let\@combinedblfloats@top\@combinedblfloats
\def\@combinedblfloats{%
  \@tempswafalse
  \let\@elt\@checkbelowtext\@dbltoplist\let\@elt\relax
  \if@tempswa
    \setbox\@tempboxa\vbox{}%
    \let\@elt\@comdblflelt\@dbltoplist\let\@elt\relax
    \xdef\@freelist{\@freelist\@dbltoplist}%
    \global\let\@dbltoplist\@empty
    \dimen@\dp\@outputbox
    \setbox\@outputbox\vbox to\textheight{%
      \unvbox\@outputbox\kern-\dimen@
      \vskip\dbltextfloatsep
      \unvbox\@tempboxa\vskip-\dblfloatsep}%
  \else
    \@combinedblfloats@top
  \fi}
\makeatother

\title{LiveAudio: From Bidirectional to Autoregressive Diffusion for Real-Time Streaming Video-to-Audio Generation}

\author{
Yuxuan Jiang$^{1}$ \quad Zehua Chen$^{1}$ \quad Yusheng Dai$^{2}$ \quad Andong Wang$^{3}$ \quad Weibei Dou$^{1}$ \quad Jun Zhu$^{1,\dagger}$\\[2pt]
$^{1}$Tsinghua University \quad $^{2}$Monash University \quad $^{3}$Renmin University of China
}

\begin{document}
\maketitle
\begingroup\renewcommand\thefootnote{\fnsymbol{footnote}}\footnotetext[2]{Corresponding author.}\endgroup

\begin{abstract}
Video-to-audio generation synthesizes sound that matches a video in content and timing, with applications ranging from film post-production to live streaming, games and interactive world models. Interactive applications receive video continuously and need the sound to follow it in real time, yet most existing video-to-audio models are bidirectional and must observe an entire clip before producing any sound. We propose LiveAudio, a streaming video-to-audio framework that generates high-quality and synchronized audio in real time while the video is playing. In contrast to prior streaming models trained from scratch, LiveAudio converts a bidirectional model into a streaming generator so as to build on its audio quality and synchronization. LiveAudio generates audio block by block, and each block depends only on the video that has already been played. The streaming generator is obtained from the bidirectional model through progressive causal fine-tuning and distillation, so that each block needs only a few sampling steps. Experiments on VGGSound show that LiveAudio outperforms existing streaming models in audio quality and synchronization, and surpasses the bidirectional model it starts from in synchronization. On a single A100 GPU, it runs in real time with a real-time factor of 0.79 and a time to first frame of 143\,ms. Audio samples are available at \url{https://streamaudio.github.io/}.
\end{abstract}

\section{Introduction}

\begin{figure}[t]
    \centering
    \includegraphics[width=\linewidth]{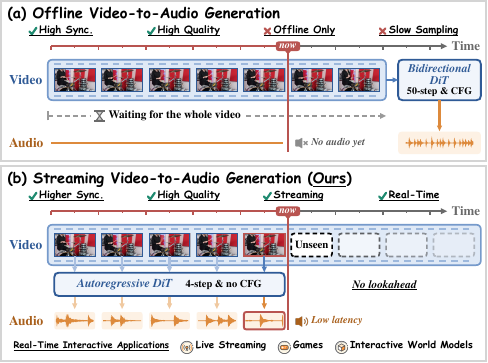}
    \caption{Comparison between (a) offline V2A generation and (b) streaming V2A generation with LiveAudio at the same moment.}
    \label{fig:teaser}
    \vspace{-2em}
\end{figure}

Video-to-audio (V2A) generation aims to synthesize sound that matches a video in both content and timing.
Recent bidirectional diffusion transformers (DiTs)~\cite{cheng2025mmaudio,liu2025thinksound,shan2025hunyuanvideo,dai2026omni2sound} have substantially improved both audio quality and audio-visual synchronization.
These models operate offline, since every audio position attends to the entire video and audio clip, and generation can start only after the whole video is available.
However, emerging interactive applications, such as live streaming~\cite{huang2026live,sun2026streamavatar}, games~\cite{che2025gamegen} and interactive world models~\cite{bruce2024genie,he2025matrix}, deliver video frame by frame and require the sound to follow it in real time.
As illustrated in Fig.~\ref{fig:teaser}, this setting calls for real-time streaming V2A generation, in which the audio depends only on the video played so far and is generated at least as fast as it is played.

Existing streaming V2A models address this setting by training a causal model from scratch.
As an early attempt, V-AURA~\cite{viertola2025temporally} generates discrete audio tokens autoregressively, conditioned on video features that are extracted segment by segment.
More recently, SoundReactor~\cite{saito2025soundreactor}, SwanSphere~\cite{lei2026towards}, WorldSonus~\cite{fang2026worldsonus} and LiveFoley~\cite{dai2026livefoley} pair a causal autoregressive backbone with a local diffusion head and generate audio as video frames arrive.
Among them, LiveFoley achieves synchronization on par with offline models~\cite{cheng2025mmaudio,liu2025thinksound,shan2025hunyuanvideo,dai2026omni2sound}, and its sync-aligned video conditioning opens a promising direction for streaming V2A generation.
Such models, however, cannot build on the audio quality and synchronization that bidirectional models have already achieved.
Both have to be learned anew, often with additional training objectives.

Video generation has taken a different route.
A line of work converts a bidirectional diffusion model into a causal, few-step autoregressive generator through causal fine-tuning and distillation~\cite{yin2025slow,huang2026self,zhu2026causal,shin2026motionstream}, and audio-driven avatar systems~\cite{huang2026live,sun2026streamavatar,ki2026avatar,zhang2026vidu} and joint audio-video generators~\cite{su2026omniforcing,ke2026helixworld} have brought this route to real-time streaming.
This route allows a streaming generator to build on a bidirectional model, but it has not been explored for V2A generation, where it faces three challenges.
First, the video condition arrives over time, so the generator may use only the frames that have been played, whereas a bidirectional V2A model receives video features that cover later frames.
Second, a causal generator cannot observe later audio or video, and we find that this degrades audio-visual synchronization.
Third, audio must be generated in real time, and the incoming video has to be encoded within the same time budget, whereas sampling alone takes tens of steps~\cite{song2020denoising,lu2025dpm,liu2022flow} with classifier-free guidance (CFG)~\cite{ho2022classifier}.

To address these challenges, we present LiveAudio, a real-time streaming V2A framework that decouples the problem into causality and efficiency, and solves them progressively starting from a bidirectional V2A model.
For causality, LiveAudio is formulated as an autoregressive diffusion model with a block-causal DiT~\cite{peebles2023scalable} and a causal audio VAE~\cite{saito2025soundreactor}, in which audio is generated block by block and each block is decoded and played immediately.
The video conditions~\cite{iashin2024synchformer} are also made causal through sync-aligned video conditioning~\cite{dai2026livefoley}, in which each audio token receives only the video that has been played.
We train a bidirectional model under this conditioning and then make it causal by block-causal fine-tuning.
This gives a causal V2A model that generates audio as the video arrives, but each block still requires multi-step sampling with CFG.
For efficiency, we distill the causal model by ODE regression~\cite{yin2025slow,huang2026self,zhu2026causal} into a few-step generator that needs no CFG and runs in real time.
Finally, we apply asymmetric distribution matching distillation (DMD)~\cite{yin2025slow,zhu2026causal}, in which the bidirectional model supervises the audio that the causal generator produces from its own history, so that synchronization is recovered without losing causality or efficiency.

We evaluate LiveAudio on the VGGSound test set~\cite{chen2020vggsound}, where it outperforms existing streaming V2A models in both audio quality and synchronization.
Its synchronization surpasses that of the bidirectional model it starts from and of the offline models we compare with, while its audio quality remains competitive.
In terms of efficiency, LiveAudio runs in real time on a single A100 GPU, with a time to first frame of 143\,ms and a real-time factor of 0.79.
Ablation studies show that every stage of the pipeline is necessary, and the same model also supports streaming video-text-to-audio (VT2A) and text-to-audio (T2A) generation.
Our contributions are summarized as follows:
\begin{itemize}
    \item We present LiveAudio, a real-time streaming V2A framework that converts a bidirectional V2A model into an autoregressive diffusion generator, in contrast to existing streaming models that are trained from scratch.
    \item We decouple this conversion into causality and efficiency with a progressive training pipeline, and show that asymmetric DMD recovers the synchronization lost to causality and improves it beyond that of the bidirectional model.
    \item LiveAudio outperforms existing streaming V2A models in audio quality and synchronization on VGGSound, runs in real time on a single GPU, and is also validated on streaming VT2A and T2A generation.
\end{itemize}

\section{Related Work}

\subsection{Video-to-Audio Generation}
Prior work explored V2A generation with latent diffusion models, either conditioned on contrastively learned audio-visual features~\cite{luo2023diff} or built by adding visual conditions to T2A models~\cite{jeong2025read,zhang2026foleycrafter,chen2026visual,dai2026cinedub}.
Recent bidirectional DiTs~\cite{cheng2025mmaudio,liu2025thinksound,shan2025hunyuanvideo,dai2026omni2sound} are trained jointly with T2A data and support video, text and joint conditioning in a single model.
Their timing is provided by video features injected in temporal alignment with the audio, typically synchronization features from Synchformer~\cite{iashin2024synchformer}, whereas semantics come from global video features and text.
All of them are offline, since audio and video are mutually visible over the entire clip.
Streaming models, from V-AURA~\cite{viertola2025temporally} to SoundReactor~\cite{saito2025soundreactor}, SwanSphere~\cite{lei2026towards}, WorldSonus~\cite{fang2026worldsonus} and LiveFoley~\cite{dai2026livefoley}, instead train a causal model from scratch, so their audio quality and synchronization are learned anew rather than inherited from bidirectional models.

\subsection{Streaming Multimodal Generation}
Real-time interactive applications require streaming generation, where the model produces output as its input arrives.
In video generation, building on Diffusion Forcing~\cite{chen2024diffusion}, a line of work has established how to turn a bidirectional diffusion model into a causal, few-step autoregressive generator~\cite{yin2025slow,huang2026self,zhu2026causal,shin2026motionstream}.
This route has been extended to settings where the condition itself arrives over time, including audio-driven avatar generation~\cite{huang2026live,sun2026streamavatar,ki2026avatar,zhang2026vidu}, and to joint audio-video generation~\cite{su2026omniforcing,ke2026helixworld}.
However, it has not been explored for V2A generation, which this work addresses.

\begin{figure*}[t]
    \centering
    \includegraphics[width=\linewidth]{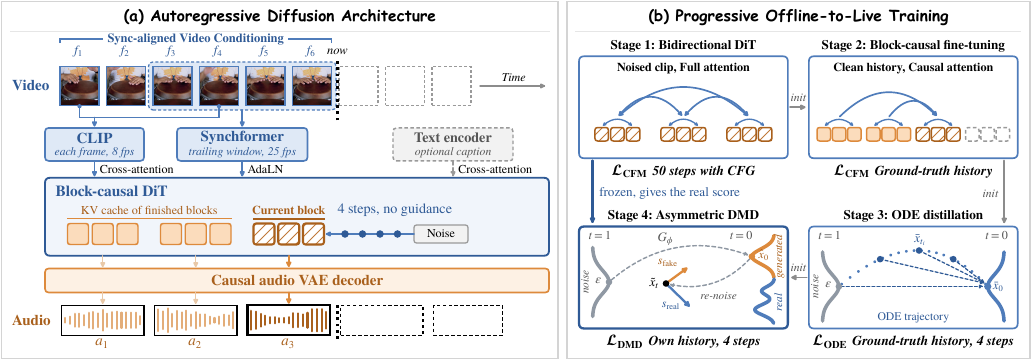}
    \caption{Overview of LiveAudio. (a) Audio is generated block by block and decoded as soon as it is generated, and sync-aligned video conditioning makes each block depend only on the video played so far. (b) Four progressive training stages turn a bidirectional model into a four-step autoregressive generator without guidance, first making it causal and then distilling it.}
    \label{fig:overview}
    \vspace{-1em}
\end{figure*}

\subsection{Few-Step Diffusion Distillation}
Few-step diffusion models are mainly trained with trajectory-based methods, which learn to reach the endpoint of the sampling trajectory in one or a few steps~\cite{song2023consistency,geng2026mean}, or with DMD~\cite{yin2024one,yin2024improved,liu2026decoupled}, which pulls the output distribution of the student toward that of the teacher without imitating its trajectory.
In audio generation, trajectory-based methods reach one- to few-step sampling in T2A~\cite{bai2023consistencytta,liu2024audiolcm,liu2025flashaudio,li2026meanaudio} and V2A~\cite{hadjeres2026woosh}, and DMD has been applied to music~\cite{novack2025presto}, sound~\cite{tian2026audiox} and speech~\cite{li2024dmdspeech}, usually with an adversarial term on real audio that improves perceptual quality~\cite{novack2025presto,tian2026audiox}.
Distilled students can even surpass their teachers~\cite{yin2024improved,li2024dmdspeech}.
Adversarial post-training can also yield few-step models from real data without a teacher~\cite{novack2025fast}, and extends to block-causal streaming music generation~\cite{novack2026live}.
However, in these methods the teacher and the student observe the same context, or no teacher is used.
Asymmetric distillation, in which a bidirectional teacher supervises a causal student, has been established for streaming video generation~\cite{yin2025slow,zhu2026causal}, but its effect on audio-visual synchronization has not been studied.

\section{LiveAudio}

\subsection{Problem Setup}
\paragraph{Real-time streaming video-to-audio generation.}
Let $y$ denote the text condition, $f_{1:T}$ the video frames, and $x_{1:K}$ the audio latents on the same timeline, divided into $K$ blocks with block $x_k$ covering $[s_k, e_k)$.
V2A generation models the conditional distribution $p(x_{1:K}\,|\, f_{1:T}, y)$, which offline models realize with bidirectional attention, so every audio position observes all video frames and the full audio context.
Real-time streaming adds two requirements.
The first is causality.
Audio is produced block by block, and block $k$ is generated once the video has played up to time $\tau_k$.
It may depend only on the frames $f_{\le\tau_k}$ that have arrived, on the text and on the previously generated audio.
Every latent of the block starts before $\tau_k$, so sound never precedes its video.
A streaming generator therefore has the form:
{\abovedisplayskip=4pt\belowdisplayskip=4pt
\begin{equation}
    p_\theta(x_{1:K}\,|\, f_{1:T}, y)=\prod_{k=1}^{K} p_\theta(x_k\,|\, c_k),
    \label{eq:streaming_factorization}
\end{equation}}%
where $c_k=(y, x_{<k}, f_{\le\tau_k})$ is the context of block $k$.
The second is real-time operation.
The computation of each block must not exceed the block duration, so that the delay of a block stays at its video wait $\tau_k-s_k$ plus its computation and does not accumulate over time.
\paragraph{Conditional flow matching.}
Flow matching learns a velocity field that transports Gaussian noise to data~\cite{lipman2022flow,liu2022flow,hung2024tangoflux,jiang2026freesonic}.
The bidirectional and causal models are trained with its conditional form, the former on a whole clip and the latter on a block $x_k$ with the context $c_k$.
For a clean sample $x$ and noise $\epsilon\sim\mathcal{N}(0,I)$, the linear path is:
{\abovedisplayskip=7pt\belowdisplayskip=7pt
\begin{equation}
    x_{t}=(1-t)\,x+t\,\epsilon,\qquad t\in[0,1],
    \label{eq:flow_path}
\end{equation}}%
where larger $t$ means more noise, and a velocity model $v_\theta$ is trained by regressing the velocity of this path~\cite{cheng2025mmaudio}:
{\abovedisplayskip=7pt\belowdisplayskip=7pt
\begin{equation}
    \mathcal{L}_{\mathrm{CFM}}
    =\mathbb{E}_{t,\,x,\,\epsilon}
    \left[\left\|v_\theta(x_{t},c,t)-(\epsilon-x)\right\|_2^2\right].
    \label{eq:conditional_fm}
\end{equation}}%
For CFG~\cite{ho2022classifier}, the video and text in $c$ are randomly replaced by empty conditions during training, giving the context $\varnothing$.
At inference, a sample is obtained from $\epsilon$ by solving the following ODE from $t=1$ to $t=0$ with guidance scale $w$:
{\abovedisplayskip=7pt\belowdisplayskip=7pt
\begin{equation}
    \frac{\mathrm{d}x_{t}}{\mathrm{d}t}=v_\theta(x_{t},\varnothing,t)+w\,\big[v_\theta(x_{t},c,t)-v_\theta(x_{t},\varnothing,t)\big].
    \label{eq:sampling_ode}
\end{equation}}%
\paragraph{Distribution matching distillation.}
DMD~\cite{yin2024one,yin2024improved} distills a multi-step teacher into a few-step generator by matching output distributions rather than sampling trajectories.
The generator $G_\phi$ maps noise $\epsilon$ and context $c$ to a sample $\hat{x}=G_\phi(\epsilon,c)$.
Re-noising $\hat{x}$ to noise level $t$ by Eq.~\ref{eq:flow_path} with fresh noise $\epsilon'$ gives $\tilde{x}_{t}$.
DMD minimizes the expected reverse KL divergence between the distributions of the generator and the teacher at noise level $t$, with gradient:
{\abovedisplayskip=7pt\belowdisplayskip=7pt
\begin{equation}
    \nabla_\phi\mathcal{L}_{\mathrm{DMD}}
    =\mathbb{E}_{t,\,\epsilon,\,\epsilon'}\!\left[
        \big(s_{\mathrm{fake}}-s_{\mathrm{real}}\big)(\tilde{x}_{t},c,t)\,
        \frac{\partial\tilde{x}_{t}}{\partial\phi}
    \right],
    \label{eq:dmd_grad}
\end{equation}}%
where $s_{\mathrm{real}}$ and $s_{\mathrm{fake}}$ are the scores of the two distributions, called the real score and the fake score.
The teacher is frozen and gives the real score with the guidance of Eq.~\ref{eq:sampling_ode}, while a fake score model, trained online on the generator's samples with Eq.~\ref{eq:conditional_fm}, gives the fake score.
Both scores are computed from the velocity predictions of the two models.
\subsection{Autoregressive Diffusion Architecture}
\label{sec:architecture}
Fig.~\ref{fig:overview} (a) illustrates the autoregressive diffusion architecture of LiveAudio.
To satisfy the causality requirement, audio is encoded by a causal audio VAE and generated by a block-causal DiT, so that each block can be decoded and played as soon as it is generated.
The video conditions are also made causal by injecting them in a sync-aligned manner, so that each block depends only on the video played so far.

\paragraph{Causal audio VAE.}
The audio VAEs of existing T2A and V2A systems use non-causal convolutions~\cite{liu2023audioldm,evans2025stable,cheng2025mmaudio,dai2026omni2sound}, so decoding a latent requires the latents that follow it.
Thus, a block can be decoded faithfully only after later blocks are generated, and decoding it earlier produces boundary artifacts.
We instead adopt the causal audio VAE of SoundReactor~\cite{saito2025soundreactor} and keep it frozen throughout training.
It encodes 48~kHz stereo audio into 64-dimensional latents at 30~Hz.
Its encoder and stateful decoder depend only on past context, so a block of latents is decoded to waveform as soon as it is generated, without re-decoding earlier blocks.
\paragraph{Block-causal DiT.}
We use a diffusion transformer~\cite{peebles2023scalable} as the backbone and support streaming without adding parameters.
Each latent is one audio token, and we group every three consecutive tokens into a block of 0.1~s.
The self-attention mask is block-causal~\cite{yin2025slow,huang2026self}, so that a token attends to all tokens of its own block and of earlier blocks, but not to those of later blocks.
The bidirectional model prepends one token that carries the noise level of the whole clip, and we give every block its own such token, so that blocks can be at different noise levels.
During training, the current block is noised while earlier blocks stay clean.
At inference, audio is generated block by block under the same mask, and finished blocks serve as the condition of later ones through a KV cache.
A single block that spans the clip recovers the bidirectional model, so weights transfer between the two without any architectural change.
\paragraph{Sync-aligned video conditioning.}
We keep the two video paths of MMAudio~\cite{cheng2025mmaudio} and Omni2Sound~\cite{dai2026omni2sound}, in which Synchformer~\cite{iashin2024synchformer} features modulate each audio token through adaptive layer normalization and CLIP~\cite{radford2021learning} features enter through attention together with the text.
For streaming, we change how the two paths are fed, so that a token receives only video that has been played.
We call two consecutive video frames a slice.
For Synchformer, we adopt the trailing sliding window of LiveFoley~\cite{dai2026livefoley}, which encodes the most recent frames each time a slice has been played, and each audio token takes the feature of the slice that contains its start.
For CLIP, which is sampled at a lower frame rate, the cross-attention of a token covers only the frames that have arrived by the end of its slice, while text is visible throughout.
Block $k$ is generated at the time $\tau_k$ when the slices of all its tokens have been played.
A block therefore waits only for the slice of its last token and uses no further lookahead, so the video wait $\tau_k-s_k$ stays close to the duration of one block.
Timing cues are thus local and arrive together with the frames of each block, whereas semantic context accumulates as the video plays.

\begin{table*}[t]
    \centering
    \vspace{-0.5em}
    \small
    \caption{Video-to-audio results on the VGGSound test set. Offline models generate only after the whole video is available, so TTFF is reported only for streaming models. Results marked with $\dagger$ are quoted from the original papers. The teacher is our bidirectional DiT.}
    \setlength{\tabcolsep}{5.05pt}
    \begin{tabular}{l ccccc cc ccc cc}
        \toprule
        \multirow{2}{*}{Method} & \multicolumn{5}{c}{Distribution Matching} & \multicolumn{2}{c}{Audio Quality} & \multicolumn{3}{c}{Modality Alignment} & \multicolumn{2}{c}{Efficiency} \\
        \cmidrule(lr){2-6}\cmidrule(lr){7-8}\cmidrule(lr){9-11}\cmidrule(lr){12-13}
        & FD$_{\mathrm{O}}$$\downarrow$ & FD$_{\mathrm{V}}$$\downarrow$ & FD$_{\mathrm{P}}$$\downarrow$ & KL$_{\mathrm{S}}$$\downarrow$ & KL$_{\mathrm{P}}$$\downarrow$ & IS$_{\mathrm{P}}$$\uparrow$ & PQ$\uparrow$ & DeSync$\downarrow$ & IB$\uparrow$ & CLAP$\uparrow$ & TTFF (s)$\downarrow$ & RTF$\downarrow$ \\
        \midrule
        \multicolumn{13}{l}{\textit{Offline}} \\
        FRIEREN & 222.58 & 1.35 & 11.73 & 2.95 & 2.74 & 12.12 & 5.90 & 0.849 & 0.228 & 0.227 & - & 0.13 \\
        MMAudio & 65.85 & 0.76 & 4.57 & 1.93 & 2.10 & 14.20 & 5.90 & 0.457 & 0.327 & 0.308 & - & 0.37 \\
        AudioX & 77.75 & 1.56 & 12.51 & 3.14 & 2.93 & 13.68 & 6.13 & 1.213 & 0.250 & 0.215 & - & 1.37 \\
        Omni2Sound & 143.81 & 0.63 & 6.23 & 2.11 & 2.16 & 16.70 & 6.08 & 0.501 & 0.338 & 0.284 & - & 1.06 \\
        \rowcolor{oursrow} LiveAudio (teacher) & 68.17 & 1.29 & 5.11 & 1.86 & 1.99 & 15.81 & 6.26 & 0.484 & 0.324 & 0.260 & - & 0.50 \\
        \midrule
        \multicolumn{13}{l}{\textit{Streaming}} \\
        V-AURA & 82.76 & 2.88 & 14.71 & 2.24 & 2.46 & 10.10 & 6.00 & 0.678 & 0.279 & 0.252 & 7.148 & 2.31 \\
        SoundReactor$^\dagger$ & - & - & - & 2.37 & - & - & - & 1.030 & 0.240 & - & - & - \\
        WorldSonus & 90.24 & 1.67 & 15.59 & 2.44 & 2.54 & 9.09 & 6.28 & 0.659 & 0.258 & 0.197 & 0.502 & 0.60 \\
        LiveFoley$^\dagger$ & - & - & 9.92 & - & 2.30 & 11.42 & 6.26 & 0.480 & 0.290 & - & 0.156 & - \\
        \rowcolor{oursrow} LiveAudio (ours) & 70.97 & 1.30 & 7.32 & 1.99 & 2.12 & 15.87 & 6.41 & 0.395 & 0.323 & 0.252 & 0.143 & 0.79 \\
        \bottomrule
    \end{tabular}
    \vspace{-0.5em}
    \label{tab:main}
\end{table*}

\subsection{Progressive Offline-to-Live Training}
An offline bidirectional model produces high-quality and well-synchronized audio, but it is neither causal nor real-time.
As shown in Fig.~\ref{fig:overview} (b), we turn it into a real-time streaming generator progressively~\cite{jiang2026controlaudio} in four stages, where each stage after the first starts from the weights of the previous one and changes only one property.
We first train the bidirectional model as the starting point, then make it causal, then distill the causal model into a few-step generator that runs in real time, and finally recover its synchronization with asymmetric DMD.
\paragraph{Stage 1: Bidirectional DiT pretraining.}
We train a bidirectional DiT~\cite{evans2025stable} with flow matching (Eq.~\ref{eq:conditional_fm}) on video-audio, video-text-audio and text-audio data in the latent space of the causal VAE~\cite{saito2025soundreactor}.
All audio tokens of a clip attend to each other bidirectionally.
In MMAudio~\cite{cheng2025mmaudio} and Omni2Sound~\cite{dai2026omni2sound}, the Synchformer and CLIP features that an audio token receives cover later frames.
We instead adopt sync-aligned video conditioning for both, so a token receives only video that has been played.
Our bidirectional DiT is therefore bidirectional only in the attention among audio tokens, while its video conditioning is already causal and remains unchanged in all later stages.
\paragraph{Stage 2: Block-causal fine-tuning.}
We make the generation of audio causal in this stage and change nothing else.
We switch the bidirectional DiT to the block-causal structure and fine-tune it with the same objective, data and video conditioning.
Block $k$ now observes only the context $c_k$ of Eq.~\ref{eq:streaming_factorization}, with the earlier blocks but not the later ones.
Training uses teacher forcing~\cite{zhu2026causal}, in which the history $x_{<k}$ is the clean ground truth and only the block itself is noised.
Each block draws its own noise level, and all blocks of a clip are supervised in parallel.
The resulting causal DiT generates audio block by block but is still a multi-step flow model.
\paragraph{Stage 3: Causal ODE distillation for efficiency.}
The causal DiT is still not real-time, since every block needs many sampling steps and each step evaluates the model twice for CFG.
We therefore adopt ODE regression~\cite{yin2025slow,huang2026self,zhu2026causal} to distill it into a few-step generator $G_\phi$ that needs no guidance.
Solving Eq.~\ref{eq:sampling_ode} with the causal DiT gives a deterministic trajectory from noise to a clean block, and the guidance is already contained in this trajectory.
We extract such trajectories with ground-truth history and store their states $\bar{x}_{k,t_i}$ at the step times $t_i$ of the few-step sampler together with their endpoints $\bar{x}_{k,0}$.
The generator is initialized from the causal DiT and trained to predict the endpoint from the stored state of a random step:
\begin{equation}
    \mathcal{L}_{\mathrm{ODE}}
    =\mathbb{E}_{\bar{x},\,k,\,i}\!\left[\left\|G_\phi(\bar{x}_{k,t_i},c_k,t_i)-\bar{x}_{k,0}\right\|_2^2\right],
    \label{eq:ode_regression}
\end{equation}
where $G_\phi(\cdot,c_k,t_i)$ is one forward pass of the generator.
The generator now produces a block in a few steps without guidance, but it has never been conditioned on its own outputs.
\paragraph{Stage 4: Asymmetric DMD for synchronization.}
The ODE-distilled generator $G_\phi$ is causal and produces each block in a few steps, but its audio-visual synchronization remains inferior to that of the bidirectional DiT.
This is because a causal model cannot observe later audio or video, and the generator has so far been trained only with ground-truth history rather than its own outputs.
We therefore adopt asymmetric DMD~\cite{yin2025slow,zhu2026causal}, in which the bidirectional DiT, trained on real audio and attending over the whole clip, supervises the outputs of the causal generator.
DMD matches distributions rather than individual trajectories, so the teacher does not need to observe the same context as the generator.
The generator produces a whole clip block by block, where the history $\hat{x}_{<k}$ of block $k$ is its own output rather than the ground truth~\cite{huang2026self}.
The generated clip is re-noised, and the generator is updated with the gradient of Eq.~\ref{eq:dmd_grad}, where the real score is given by the bidirectional DiT with guidance and the fake score by a copy of it trained online on the generated clips.
Each generated clip is thus assessed as a whole by the bidirectional DiT, so the generator learns to produce well-synchronized audio while remaining causal.
We further add an adversarial loss, in which a discriminator head on the features of the fake score model separates real audio latents from generated ones~\cite{yin2024improved,novack2025presto}.

\subsection{Block-wise KV-Cached Inference}
At inference, the generator realizes the factorization of Eq.~\ref{eq:streaming_factorization} by producing the blocks sequentially.
Real-time operation requires the cost of a block to be bounded by its duration, which precludes recomputing the history $\hat{x}_{<k}$ for every new block.
Since the self-attention is block-causal, the keys and values of a finished block are independent of all subsequent blocks and are therefore retained in a KV cache~\cite{huang2026self,yin2025slow}.
Block $k$ is thus generated at time $\tau_k$ as:
\begin{equation}
    \hat{x}_k=G_\phi\big(\epsilon_k;\;y,\;f_{\le\tau_k},\;\mathrm{KV}(\hat{x}_{<k})\big),
    \label{eq:cached_inference}
\end{equation}
where $\epsilon_k\sim\mathcal{N}(0,I)$ is the initial noise and $\mathrm{KV}(\hat{x}_{<k})$ denotes the cached self-attention keys and values of the finished blocks.
The generator processes only the tokens of block $k$, which attend to the cache.
It runs the few-step sampler, which alternates between predicting the clean block and re-noising the prediction to the next step time by Eq.~\ref{eq:flow_path} with fresh noise.
The keys and values of the finished block are then obtained from one additional pass on the clean block and appended to the cache, so that the cached history remains clean.
The finished block is decoded by the causal VAE and played immediately, while the features of incoming video slices are computed incrementally.

\section{Experiments}

\subsection{Experimental Setup}
\paragraph{Datasets.}
We train all four stages on the same corpus.
The video-audio data come from SoundAtlas~\cite{dai2026omni2sound}, which contains 470k video-audio-text pairs from \mbox{VGGSound}~\cite{chen2020vggsound} and AudioSet~\cite{gemmeke2017audio}, and the text-audio data from AudioCaps~\cite{kim2019audiocaps} and WavCaps~\cite{mei2024wavcaps}.
We evaluate V2A and VT2A generation on the VGGSound test set, with VT2A text prompts from VGGSound-Omni~\cite{dai2026omni2sound}, and T2A generation on the AudioCaps test set.
More details are provided in Appendix~\ref{app:datasets}.

\paragraph{Evaluation metrics.}
We follow the evaluation protocols of \mbox{AudioLDM}~\cite{liu2023audioldm}, Stable Audio Open~\cite{evans2025stable} and MMAudio~\cite{cheng2025mmaudio}, and evaluate distribution matching, audio quality, modality alignment and efficiency.
For distribution matching, we report the Fr\'echet distance between generated and ground-truth audio on OpenL3~\cite{cramer2019look}, VGGish~\cite{hershey2017cnn} and PANNs~\cite{kong2020panns} embeddings, denoted FD$_{\mathrm{O}}$, FD$_{\mathrm{V}}$ and FD$_{\mathrm{P}}$.
We also report the KL divergence between the class distributions that PaSST~\cite{koutini2021efficient} and PANNs predict for generated and ground-truth audio, denoted KL$_{\mathrm{S}}$ and KL$_{\mathrm{P}}$.
For audio quality, we report the inception score computed with PANNs (IS$_{\mathrm{P}}$) and the production quality (PQ) predicted by Audiobox Aesthetics~\cite{tjandra2025meta}.
For modality alignment, we report the desynchronization score (DeSync), the mean absolute audio-visual offset predicted by Synchformer~\cite{iashin2024synchformer}.
We also report the ImageBind~\cite{girdhar2023imagebind} similarity between the video and the generated audio (IB) and the CLAP~\cite{wu2023large} similarity between the generated audio and the caption.
For efficiency, we report the time to first frame (TTFF), measured from the start of the video stream to the first audio output, and the real-time factor (RTF), the ratio of the processing time to the duration of the generated audio.
We further conduct a subjective evaluation with 20 participants and report mean opinion scores for overall audio quality (MOS-Q) and temporal alignment (MOS-T).
Details of all metrics are provided in Appendix~\ref{app:metrics}.

\paragraph{Implementation details.}
We use a DiT~\cite{evans2025stable} with 24 layers, 24 attention heads and a width of 1536, about 1.2B parameters, and train it on 10\,s clips.
Synchformer runs on 25\,fps video, so a slice covers 0.08\,s, and CLIP runs on 8\,fps video.
The resulting video wait of a block averages 0.11\,s.
The optional text prompt is encoded by a frozen Flan-T5 base~\cite{chung2024scaling} into at most 128 tokens.
We train the bidirectional DiT for 500k steps, fine-tune the block-causal DiT for 100k steps, train the generator by ODE regression for 30k steps on 131k extracted trajectories, and run asymmetric DMD for 3k iterations.
The bidirectional and block-causal DiTs are sampled with 50 Euler steps and a guidance scale of 4.5, whereas the generator takes four steps without guidance.
We measure TTFF and RTF on a single NVIDIA A100 GPU with a batch size of one, and the processing time covers video decoding, feature extraction, sampling and waveform decoding.
More details are provided in Appendix~\ref{app:implementation}.

\subsection{Main Results}
\begin{table}[t]
    \centering
    \small
    \caption{Subjective evaluation of video-to-audio generation. We report mean opinion scores with 95\% confidence intervals.}
    \setlength{\tabcolsep}{4pt}
    \begin{tabularx}{\linewidth}{l *{2}{>{\centering\arraybackslash}X}}
        \toprule
        Method & MOS-Q$\uparrow$ & MOS-T$\uparrow$ \\
        \midrule
        Ground Truth & 4.06$_{\pm0.25}$ & 4.26$_{\pm0.17}$ \\
        \midrule
        \multicolumn{3}{l}{\textit{Offline}} \\
        FRIEREN & 3.28$_{\pm0.21}$ & 3.47$_{\pm0.22}$ \\
        MMAudio & 3.80$_{\pm0.28}$ & 3.69$_{\pm0.20}$ \\
        AudioX & 3.16$_{\pm0.21}$ & 2.13$_{\pm0.23}$ \\
        Omni2Sound & 3.87$_{\pm0.26}$ & 3.97$_{\pm0.19}$ \\
        \midrule
        \multicolumn{3}{l}{\textit{Streaming}} \\
        V-AURA & 3.43$_{\pm0.24}$ & 3.50$_{\pm0.18}$ \\
        WorldSonus & 3.35$_{\pm0.25}$ & 3.42$_{\pm0.24}$ \\
        \rowcolor{oursrow} LiveAudio (ours) & 3.66$_{\pm0.19}$ & 4.06$_{\pm0.27}$ \\
        \bottomrule
    \end{tabularx}
    \vspace{-1em}
    \label{tab:subjective}
\end{table}

\begin{figure*}[t]
    \centering
    \includegraphics[width=\linewidth]{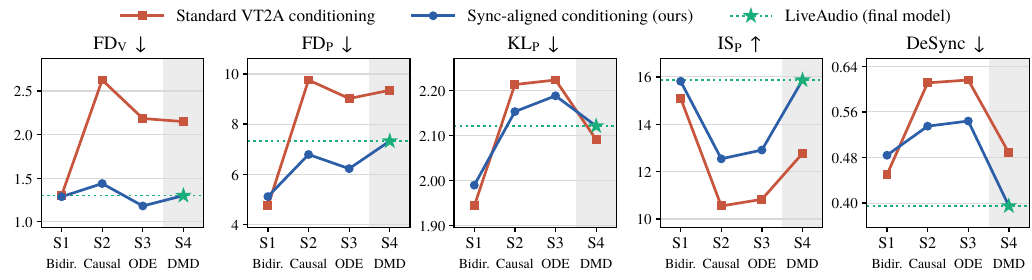}
    \caption{Ablation of the video conditioning on the VGGSound test set. Sync-aligned conditioning is compared with the standard VT2A conditioning after each of the four training stages.}
    \label{fig:conditioning}
    \vspace{-1em}
\end{figure*}

\paragraph{Video-to-audio generation.}
Our comparison covers representative offline models, namely FRIEREN~\cite{wang2024frieren}, MMAudio~\cite{cheng2025mmaudio}, AudioX~\cite{tian2025audiox} and Omni2Sound~\cite{dai2026omni2sound}, and streaming models, including V-AURA~\cite{viertola2025temporally}, SoundReactor~\cite{saito2025soundreactor} and the most recent WorldSonus~\cite{fang2026worldsonus} and LiveFoley~\cite{dai2026livefoley}.
We quote the results of SoundReactor and LiveFoley from their papers, and run all other models with their official checkpoints and inference settings.
We also report our bidirectional DiT as LiveAudio (teacher).
As shown in Table~\ref{tab:main}, LiveAudio achieves the lowest DeSync among all models and surpasses the streaming models, the offline models and its own teacher by a clear margin.
In semantic alignment, LiveAudio obtains the highest IB among the streaming models, close to that of the offline models, and its CLAP is on par with the best streaming model.
In distribution matching and audio quality, LiveAudio consistently outperforms the streaming models and is comparable to the offline models, even though it generates each block in four steps from past context only.
For efficiency, LiveAudio runs faster than real time, and its TTFF is among the lowest of the streaming models.
The subjective evaluation in Table~\ref{tab:subjective} is consistent with the objective metrics.
LiveAudio obtains the highest MOS-T among all models, and its MOS-Q is the highest among the streaming models and close to that of the best offline models.
These results show that a streaming generator converted from a bidirectional model outperforms those trained from scratch.
We attribute the advantage in synchronization to asymmetric DMD, in which the bidirectional DiT supervises the audio that the causal generator produces from its own history.
LiveAudio also supports VT2A generation with the same model, where a text prompt is provided together with the video to specify the desired sound.
The complete results are provided in Appendix~\ref{app:vt2a}.
\paragraph{Text-to-audio generation.}
We further compare LiveAudio with representative offline T2A models, namely AudioLDM~\cite{liu2023audioldm}, Tango~\cite{ghosal2023text}, AudioLDM~2~\cite{liu2024audioldm}, Stable Audio Open~\cite{evans2025stable}, TangoFlux~\cite{hung2024tangoflux}, MMAudio~\cite{cheng2025mmaudio} and AudioX~\cite{tian2025audiox}, on the AudioCaps test set~\cite{kim2019audiocaps} in Table~\ref{tab:t2a}.
We also include AudioGen~\cite{kreuk2022audiogen}, an autoregressive model that generates discrete audio tokens, and WorldSonus~\cite{fang2026worldsonus}, a streaming model.
LiveAudio is trained on text-audio pairs in all four stages, so the same model also supports streaming T2A generation without video input, in which audio is generated block by block in four steps.
Despite this causal and few-step generation, LiveAudio is comparable to the offline T2A models, with the lowest FD$_{\mathrm{V}}$ among all models, outperforms the streaming WorldSonus on all metrics, and is competitive with the autoregressive AudioGen.
More T2A results are provided in Appendix~\ref{app:t2a}.

\begin{table}[t]
    \centering
    \small
    \caption{Text-to-audio results on the AudioCaps test set. LiveAudio is the same model as in Table~\ref{tab:main} with text input only.}
    \setlength{\tabcolsep}{2.5pt}
    \begin{tabularx}{\linewidth}{X cccccc}
        \toprule
        Method & FD$_{\mathrm{V}}$$\downarrow$ & FD$_{\mathrm{P}}$$\downarrow$ & KL$_{\mathrm{P}}$$\downarrow$ & IS$_{\mathrm{P}}$$\uparrow$ & PQ$\uparrow$ & CLAP$\uparrow$ \\
        \midrule
        \multicolumn{7}{l}{\textit{Offline}} \\
        AudioLDM & 4.87 & 35.88 & 2.10 & 6.60 & 5.68 & 0.238 \\
        Tango & 4.00 & 29.75 & 1.61 & 5.08 & 5.74 & 0.330 \\
        AudioLDM 2 & 2.41 & 33.73 & 1.59 & 7.26 & 5.87 & 0.421 \\
        Stable Audio Open & 3.66 & 38.02 & 2.24 & 9.48 & 6.23 & 0.289 \\
        TangoFlux & 2.21 & 21.66 & 1.25 & 11.92 & 5.89 & 0.544 \\
        MMAudio & 3.32 & 18.33 & 1.44 & 11.52 & 5.61 & 0.489 \\
        AudioX & 2.29 & 18.70 & 1.51 & 10.76 & 5.79 & 0.476 \\
        \midrule
        \multicolumn{7}{l}{\textit{Streaming}} \\
        AudioGen & 2.11 & 17.51 & 1.54 & 9.34 & 5.28 & 0.455 \\
        WorldSonus & 2.46 & 20.65 & 1.66 & 8.63 & 5.81 & 0.442 \\
        \rowcolor{oursrow} LiveAudio (ours) & 2.02 & 19.12 & 1.65 & 10.76 & 6.09 & 0.460 \\
        \bottomrule
    \end{tabularx}
    \vspace{-1em}
    \label{tab:t2a}
\end{table}

\subsection{Ablation Studies and Analysis}
\label{sec:ablation}

\paragraph{Ablation of progressive training.}
We analyze the progressive training in Table~\ref{tab:ablation}, where (a) reports the results after each of the four stages and (b) reports variants that skip stages before Stage 4 with the same training budget.
In (a), block-causal fine-tuning makes the model causal but lowers its audio quality and synchronization.
ODE distillation reduces sampling to four steps without guidance and preserves the quality of the block-causal DiT, while the synchronization does not improve.
Asymmetric DMD then lowers DeSync from 0.544 to 0.395, so the synchronization lost to causality is not only recovered but improved beyond the 0.484 of the bidirectional DiT within the streaming architecture.
Stage 4 slightly increases FD$_{\mathrm{V}}$ and FD$_{\mathrm{P}}$ compared with Stage 3, which is consistent with the mode-seeking behavior of DMD~\cite{yin2024one,yin2024improved}.
As in prior work~\cite{novack2025presto,tian2026audiox}, its adversarial loss mainly improves audio quality (Appendix~\ref{app:dmd}).
In (b), applying Stage 4 directly to the bidirectional DiT does not yield a usable generator.
Skipping Stage 3 (1$\,\to\,$2$\,\to\,$4) applies Stage 4 directly to the block-causal DiT, and skipping Stage 2 (1$\,\to\,$3$\,\to\,$4) follows CausVid~\cite{yin2025slow} and Self Forcing~\cite{huang2026self}, which initialize the generator by ODE regression on the trajectories of the bidirectional DiT.
Both are inferior to the full pipeline on all metrics, so every stage is necessary.

\begin{figure*}[t]
    \centering
    \includegraphics[width=\linewidth]{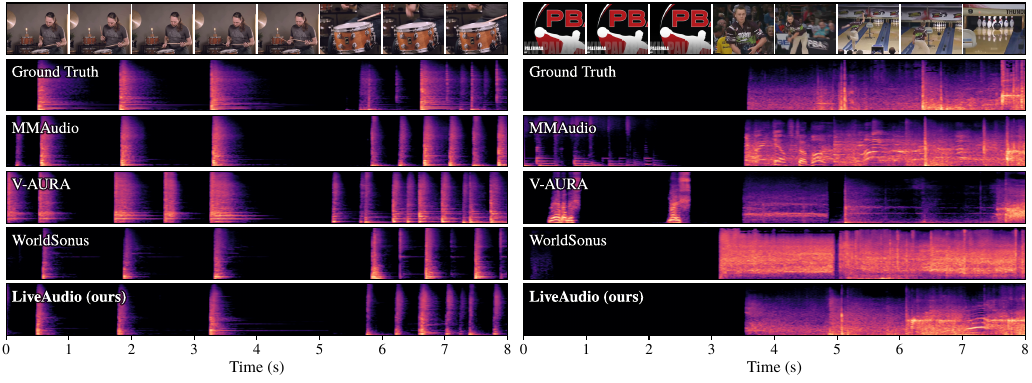}
    \caption{Qualitative comparison on two test clips, shown as mel spectrograms with the video frames placed on the same time axis. The sound events of LiveAudio are the closest to the ground truth, whereas the other models often generate sounds that are absent from it.}
    \label{fig:qualitative}
    \vspace{-1em}
\end{figure*}

\begin{table}[t]
    \centering
    \small
    \caption{Ablation of the progressive training on the VGGSound test set. (a) After each stage. (b) With some stages skipped.}
    \setlength{\tabcolsep}{1.4pt}
    \begin{tabularx}{\linewidth}{X cccccc}
        \toprule
        Stages & FD$_{\mathrm{V}}$$\downarrow$ & FD$_{\mathrm{P}}$$\downarrow$ & KL$_{\mathrm{P}}$$\downarrow$ & IS$_{\mathrm{P}}$$\uparrow$ & PQ$\uparrow$ & DeSync$\downarrow$ \\
        \midrule
        \multicolumn{7}{l}{\textit{(a) After each stage}} \\
        Stage 1 (Bidirectional) & 1.29 & 5.11 & 1.99 & 15.81 & 6.26 & 0.484 \\
        Stage 2 (Block-causal) & 1.44 & 6.79 & 2.15 & 12.54 & 6.24 & 0.535 \\
        Stage 3 (ODE) & 1.18 & 6.22 & 2.19 & 12.91 & 6.21 & 0.544 \\
        \rowcolor{oursrow} Stage 4 (DMD) & 1.30 & 7.32 & 2.12 & 15.87 & 6.41 & 0.395 \\
        \midrule
        \multicolumn{7}{l}{\textit{(b) Skipping stages}} \\
        1$\,\to\,$4 & 7.89 & 35.68 & 2.60 & 5.90 & 5.95 & 0.621 \\
        1$\,\to\,$2$\,\to\,$4 & 1.87 & 9.76 & 2.18 & 13.94 & 6.37 & 0.425 \\
        1$\,\to\,$3$\,\to\,$4 & 2.46 & 11.05 & 2.14 & 12.79 & 6.21 & 0.423 \\
        \rowcolor{oursrow} 1$\,\to\,$2$\,\to\,$3$\,\to\,$4 (ours) & \textbf{1.30} & \textbf{7.32} & \textbf{2.12} & \textbf{15.87} & \textbf{6.41} & \textbf{0.395} \\
        \bottomrule
    \end{tabularx}
    \vspace{-1em}
    \label{tab:ablation}
\end{table}
\paragraph{Comparison of video conditioning.}
We compare sync-aligned video conditioning with the standard VT2A conditioning~\cite{cheng2025mmaudio,dai2026omni2sound} in Fig.~\ref{fig:conditioning}, where the two models are trained through the same four stages with identical settings.
The two differ in lookahead.
The standard conditioning provides each audio token with the features of a whole video segment, which contains frames after the token, whereas sync-aligned conditioning uses only the frames that have been played.
In the bidirectional stage the whole video is available, and the lookahead gives the standard conditioning a slightly lower DeSync at comparable audio quality.
In streaming generation, however, the frames after the current block have not arrived, so a model trained with the standard conditioning loses the lookahead it relies on.
From Stage 2 on, it therefore falls clearly behind sync-aligned conditioning in FD$_{\mathrm{V}}$, FD$_{\mathrm{P}}$, IS$_{\mathrm{P}}$ and DeSync, and after asymmetric DMD its DeSync is 0.488 against 0.395.
These results show that a streaming generator requires a video condition without lookahead from the bidirectional stage on.

\begin{figure}[t]
    \centering
    \includegraphics[width=\linewidth]{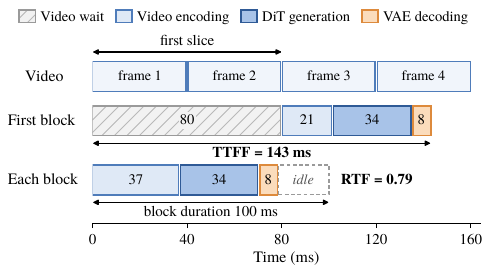}
    \caption{Latency breakdown of LiveAudio in milliseconds for the first block and for each block, with DiT generation and VAE decoding averaged over all blocks.}
    \label{fig:latency}
    \vspace{-1.5em}
\end{figure}

\paragraph{Analysis of efficiency.}
Fig.~\ref{fig:latency} breaks down the latency of LiveAudio.
For the first block, more than half of the TTFF is the wait for the first video slice, and the rest is shared by video encoding, generation and decoding.
In steady state, a block spans 1.25 video slices on average and also carries its share of the CLIP encoding, so its video encoding takes longer than that of the first block.
Each block is still processed within its duration, so LiveAudio runs faster than real time.
When the frames arrive in real time, a constant playout delay of 0.23\,s suffices for uninterrupted playback.

\paragraph{Additional studies.}
Due to the page limit, we provide further studies in Appendix~\ref{app:ablation}.
We ablate the number of sampling steps, report the numerical results of the video conditioning comparison, and examine the design of asymmetric DMD, namely its adversarial loss, its bidirectional teacher and its use of self-generated history.
We also evaluate long-form generation, where LiveAudio extends beyond its training length without additional training.

\subsection{Case Study}
Fig.~\ref{fig:qualitative} compares the generated audio on two test clips.
In the first clip, LiveAudio places the three separate strikes on the snare drum at the times of the ground truth and follows the roll, whereas V-AURA produces additional strikes.
In the second clip, the video opens with a title card, during which the ground truth is silent.
LiveAudio remains silent over this interval and starts with the scene, whereas MMAudio and V-AURA generate sound during the title card and WorldSonus starts early.
More cases are provided in Appendix~\ref{app:case_study}, and audio samples are available on the demo page.

\section{Conclusion}

We present LiveAudio, a framework that converts a bidirectional V2A model into an autoregressive generator for real-time streaming V2A generation.
LiveAudio decouples this conversion into causality and efficiency.
Sync-aligned video conditioning and block-causal fine-tuning make the model causal, and ODE distillation reduces sampling to four steps without guidance.
We show that asymmetric DMD recovers the synchronization lost when the model becomes causal and improves it beyond that of the bidirectional model.
On the VGGSound test set, LiveAudio outperforms existing streaming models in audio quality and synchronization, surpasses offline models in synchronization, and runs faster than real time on a single GPU.
Meanwhile, the same model also supports streaming VT2A and T2A generation with competitive performance.

\section*{LLM Usage Statement}
We used LLM-based assistants, Claude Code (Anthropic) and Codex (OpenAI), in the following ways.
Writing: to help draft and revise the text of all sections and to search for and summarize related work.
Figures: to help write the plotting scripts; no figure contains AI-generated images.
Code and experiments: to help write and review the training, inference and evaluation code and to launch and monitor experiments.
Ideation: the research problem, the method and the experiments to run were decided by the authors, and the assistants proposed implementation and evaluation details.
The authors reviewed all text, references, code and results, and take full responsibility for the content of the paper.

{\small
\bibliographystyle{ieeenat_fullname}
\bibliography{liveaudio}
}

\clearpage
\onecolumn
\makeatletter
\section*{Table of Contents}
{\def\app@eatdot\hskip -1em.{}%
\def\numberline#1{\hb@xt@\@tempdima{#1\hfil}\app@eatdot}%
\@starttoc{toc}}
\makeatother
\twocolumn
\appendix
\raggedbottom
\makeatletter
\setlength{\@fptop}{0pt}
\setlength{\@fpsep}{\floatsep}
\setlength{\@fpbot}{0pt plus 1fil}
\makeatother

\section{Dataset Details}
\label{app:datasets}
We train all four stages on the same corpus.
The video-to-audio (V2A) and video-text-to-audio (VT2A) tasks are trained on SoundAtlas~\cite{dai2026omni2sound}, and the text-to-audio (T2A) task on AudioCaps~\cite{kim2019audiocaps} and WavCaps~\cite{mei2024wavcaps}.
We evaluate V2A and VT2A on the VGGSound test set~\cite{chen2020vggsound}, and T2A on the AudioCaps test set.

\suppressfloats[t]
\begin{figure}[t]
    \centering
    \includegraphics[width=\linewidth]{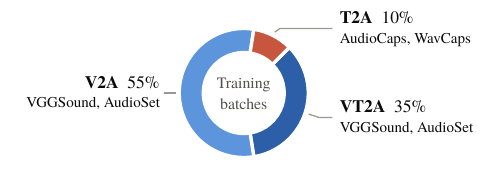}
    \caption{Probabilities of the three tasks in the training batches, which follow Omni2Sound~\cite{dai2026omni2sound}.}
    \label{fig:task_mix}
    \vspace{-1.5em}
\end{figure}

\subsection{Training Data}
\label{app:training_data}
\paragraph{SoundAtlas.}
This dataset from Omni2Sound~\cite{dai2026omni2sound} provides 470k video-audio-text triplets for clips from VGGSound~\cite{chen2020vggsound} and AudioSet~\cite{gemmeke2017audio}, two collections of 10\,s YouTube videos labelled with sound classes.
The caption of each clip describes its sound and is produced by an agentic annotation pipeline that draws on both the audio and the video of the clip.
VT2A training uses the clips with their captions, and V2A training uses all clips with the captions removed.

\paragraph{AudioCaps.}
This audio captioning dataset~\cite{kim2019audiocaps} pairs 10\,s clips from AudioSet with captions written by human annotators through crowdsourcing.
Each training clip has one caption, and each validation or test clip has five.
Its test set is a standard benchmark for T2A generation~\cite{liu2023audioldm,evans2025stable}.
We use its 50k training clips for T2A training.

\paragraph{WavCaps.}
This dataset~\cite{mei2024wavcaps} consists of 400k audio clips collected from FreeSound, BBC Sound Effects, SoundBible and AudioSet, whose captions are generated by ChatGPT from the original descriptions of the clips.
We also use it for T2A training, and divide its long recordings into segments of at most 30\,s.
Since the released audio is downsampled to 32\,kHz, we follow MMAudio~\cite{cheng2025mmaudio} and replace the clips from BBC Sound Effects with their original recordings, which are sampled at 44.1 or 48\,kHz.

\paragraph{Filtering and sampling.}
We discard clips and segments shorter than 1\,s, and remove every training clip that shares a source video with the validation or test sets of AudioCaps or with the VGGSound validation or test splits used in MMAudio~\cite{cheng2025mmaudio}.
Each training batch contains a single task, and the same mixture of tasks is used in all four stages.
Following Omni2Sound~\cite{dai2026omni2sound}, VT2A, V2A and T2A batches are sampled with probabilities of 0.35, 0.55 and 0.10, respectively (Fig.~\ref{fig:task_mix}).
All samples are 10\,s long.
A video clip starts at its beginning, a longer text-audio segment is cropped at a random position, and a shorter clip is padded with a mask.

\subsection{Evaluation Data}
\label{app:evaluation_data}
\paragraph{VGGSound test set.}
Both V2A and VT2A are evaluated on the test split of VGGSound~\cite{chen2020vggsound}, where each clip is at most 10\,s long.
The dataset is collected so that the source of the sound is visible in the video, and each clip is labelled with one sound class.
For VT2A, the text prompts are the captions of VGGSound-Omni, a benchmark introduced in Omni2Sound~\cite{dai2026omni2sound} that describes the sound of each test clip.

\paragraph{AudioCaps test set.}
T2A is evaluated on the test clips of AudioCaps~\cite{kim2019audiocaps} that remain available.
Each clip is 10\,s long and has five human-written captions, one of which is used as the text prompt.

\section{Evaluation Metrics}
\label{app:metrics}
\subsection{Distribution Matching}
\label{app:distribution}
\paragraph{FD$_{\mathbf{O}}$.}
This metric is computed with the evaluation code of Stable Audio Open~\cite{evans2025stable} and compares the distributions of generated and ground-truth audio in the embedding space of OpenL3~\cite{cramer2019look}.
All audio is peak-normalized to $-1$\,dBFS and resampled to 44.1\,kHz, with mono audio copied to both channels.
The environmental-sound version of OpenL3 extracts a 512-dimensional embedding from each channel every 0.5\,s, and the embeddings of the two channels are concatenated.
FD$_{\mathrm{O}}$ is the Fr\'echet distance between two Gaussians fitted to the embeddings of all generated audio and of all ground-truth audio.
Since it is computed at 44.1\,kHz, FD$_{\mathrm{O}}$ also reflects the frequencies above 8\,kHz, which are removed in the metrics computed at 16\,kHz.

\paragraph{FD$_{\mathbf{V}}$.}
This metric compares the distributions of generated and ground-truth audio in the embedding space of the VGGish model~\cite{hershey2017cnn}, and is computed with the evaluation code of AudioLDM~\cite{liu2023audioldm}.
It is also known as the Fr\'echet audio distance (FAD)~\cite{kilgour2018fr}.
All audio is resampled to 16\,kHz, and VGGish extracts a 128-dimensional embedding from every 0.96\,s of audio.
FD$_{\mathrm{V}}$ is the Fr\'echet distance between two Gaussians fitted to the embeddings of all generated audio and of all ground-truth audio.

\paragraph{FD$_{\mathbf{P}}$.}
This metric measures the distance between the distributions of generated and ground-truth audio in the embedding space of the PANNs CNN14 classifier~\cite{kong2020panns}, and is computed with the evaluation code of AudioLDM~\cite{liu2023audioldm}.
All audio is resampled to 16\,kHz, and each clip is mapped to a single 2048-dimensional embedding by the penultimate layer of CNN14.
FD$_{\mathrm{P}}$ is the Fr\'echet distance between two Gaussians fitted to the embeddings of all generated clips and of all ground-truth clips.

\paragraph{KL$_{\mathbf{S}}$.}
Using the PaSST classifier~\cite{koutini2021efficient}, this metric measures how well the sound events in a generated clip match those in its ground truth, and is computed with the evaluation code of Stable Audio Open.
All audio is converted to mono at 32\,kHz and peak-normalized to $-1$\,dBFS.
A clip shorter than 10\,s is zero-padded to the 10\,s input length of PaSST.
The logits predicted by PaSST over the 527 AudioSet~\cite{gemmeke2017audio} classes are converted into a class distribution by a softmax.
With $p_{\mathrm{gen}}$ and $p_{\mathrm{gt}}$ denoting the distributions of a generated clip and its ground truth, KL$_{\mathrm{S}}$ is the average of the Kullback-Leibler (KL) divergence $\mathrm{KL}(p_{\mathrm{gen}}\,\|\,p_{\mathrm{gt}})$ over all clips.

\paragraph{KL$_{\mathbf{P}}$.}
Using the PANNs CNN14 classifier~\cite{kong2020panns}, this metric measures how well the sound events in a generated clip match those in its ground truth, and is computed with the evaluation code of AudioLDM~\cite{liu2023audioldm}.
All audio is resampled to 16\,kHz, and the logits that CNN14 predicts over the 527 AudioSet classes for a whole clip are converted into a class distribution by a softmax.
With $p_{\mathrm{gen}}$ and $p_{\mathrm{gt}}$ denoting the distributions of a generated clip and its ground truth, KL$_{\mathrm{P}}$ is the average of $\mathrm{KL}(p_{\mathrm{gt}}\,\|\,p_{\mathrm{gen}})$ over all clips, with the divergence taken in the opposite direction to KL$_{\mathrm{S}}$.

\subsection{Audio Quality}
\label{app:quality}
\paragraph{IS$_{\mathbf{P}}$.}
IS$_{\mathrm{P}}$ is the inception score of the generated audio, obtained with the PANNs CNN14 classifier~\cite{kong2020panns} and the evaluation code of AudioLDM~\cite{liu2023audioldm} without using the ground truth.
All audio is resampled to 16\,kHz, and the logits that CNN14 predicts over the 527 AudioSet classes for a clip $x$ are converted into a class distribution $p(y\,|\,x)$ by a softmax.
For a set of clips, the score is the exponential of the average of $\mathrm{KL}(p(y\,|\,x)\,\|\,p(y))$, where $p(y)$ is the average of $p(y\,|\,x)$ over the set.
It is high when each clip is classified with confidence and the clips cover diverse classes.
The generated clips are shuffled with a fixed seed and split into ten subsets, and IS$_{\mathrm{P}}$ is the mean of their scores.

\paragraph{PQ.}
PQ is the production quality predicted by Audiobox Aesthetics~\cite{tjandra2025meta}, which rates audio on the four axes of production quality, production complexity, content enjoyment and content usefulness, each on a scale from 1 to 10.
Production quality concerns the technical aspects of the audio, including clarity and fidelity, dynamics, frequency content and spatialization, rather than its subjective appeal.
We use the official predictor and checkpoint, which convert the audio to mono at 16\,kHz, score it in windows of 10\,s and average the window scores weighted by their lengths.
PQ is the average score over all generated clips.

\subsection{Modality Alignment}
\label{app:alignment}
\paragraph{DeSync.}
This desynchronization score measures how far the generated audio is temporally offset from the video, and is computed with the evaluation code of MMAudio~\cite{cheng2025mmaudio}.
The first 8\,s of the video are taken at 25\,fps, and the first 8\,s of the audio are converted to mono at 16\,kHz.
Synchformer~\cite{iashin2024synchformer} predicts the audio-visual offset as one of 21 classes from $-2$ to 2\,s in steps of 0.2\,s, separately on the first 4.8\,s and the last 4.8\,s of this span.
DeSync is the absolute offset averaged over the two windows and over all clips, and clips shorter than 8\,s are excluded.

\paragraph{IB.}
IB is the ImageBind score, which measures the semantic agreement between the video and the generated audio with ImageBind~\cite{girdhar2023imagebind}, and is computed with the evaluation code of MMAudio~\cite{cheng2025mmaudio}.
Four frames are taken from the first 8\,s of the video at 0.5\,fps, and the video embedding is averaged over three spatial crops and three clips of two consecutive frames.
The audio is converted to mono at 16\,kHz and cut to its first 8\,s, and its embedding is averaged over three 2\,s clips spread over the audio.
IB is the cosine similarity between the two embeddings, averaged over the same clips as DeSync.

\paragraph{CLAP.}
CLAP measures the agreement between the generated audio and a caption, and is computed with the evaluation code of Stable Audio Open~\cite{evans2025stable}.
The generated audio is converted to mono at 48\,kHz and peak-normalized to $-1$\,dBFS, and the LAION-CLAP model~\cite{wu2023large} with feature fusion embeds the audio and the caption.
CLAP is the cosine similarity between the two embeddings, averaged over all clips.
The caption is the text prompt for VT2A and T2A, and for V2A it is the VGGSound-Omni caption of the clip, which is not given to the model.

\begin{figure*}[t]
    \centering
    \includegraphics[width=\linewidth]{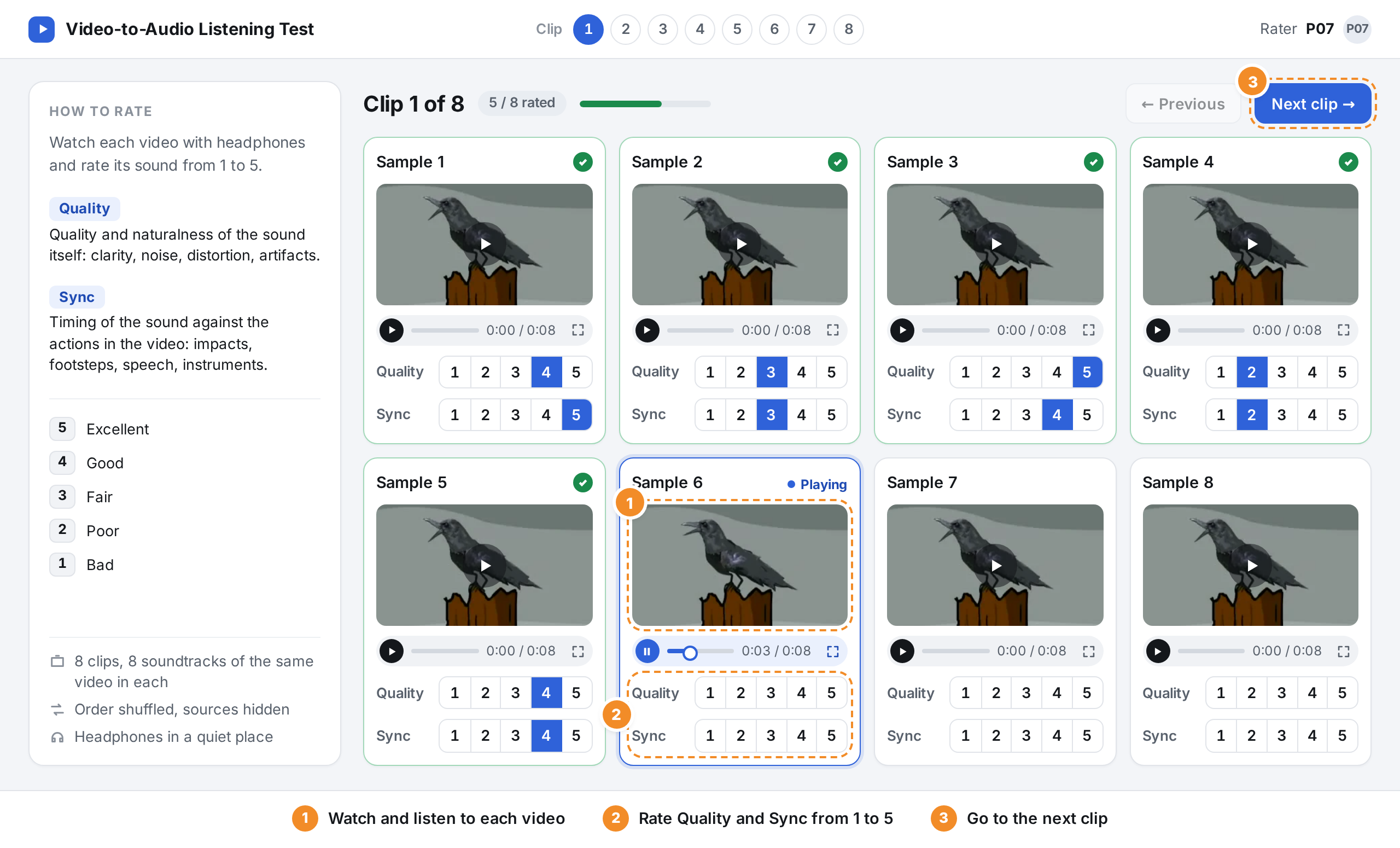}
    \caption{Interface of the subjective evaluation. Each page shows one video with its ground-truth audio and the outputs of the seven models in a random order. Participants (1) watch each video, (2) rate Quality (MOS-Q) and Sync (MOS-T) from 1 to 5, and (3) go to the next clip.}
    \label{fig:listening}
\end{figure*}

\subsection{Efficiency}
\label{app:efficiency}
Both metrics are measured on a single NVIDIA A100 GPU with a batch size of one, while no other process runs on the GPUs.
Each model uses the sampling steps, guidance scale and numerical precision of its quality evaluation.
Before timing, each model is loaded, compiled where applicable, and warmed up on four clips.
The metrics are then averaged over 20 further clips.

\paragraph{TTFF.}
The time to first frame (TTFF) is the time from the start of the video stream to the moment when the first generated audio samples are available in CPU memory.
To measure it, each video is replayed as a live stream, in which the frames are supplied already decoded and each frame becomes available only at the end of its display interval.
The streaming session of the model is opened before the stream starts.
The TTFF of a streaming model therefore includes the wait for the video that its first block requires, together with the time to encode this video, generate the block and decode it into a waveform.
Before its first block, LiveAudio waits for 0.08\,s of video, WorldSonus~\cite{fang2026worldsonus} for 0.1\,s and V-AURA~\cite{viertola2025temporally} for 2.56\,s.
An offline model produces audio only after the whole video has arrived and been processed, so its TTFF exceeds the video duration and is not reported.

\paragraph{RTF.}
RTF is the real-time factor, the ratio of the processing time of a video to the duration of the generated audio, and a value below one means that the model runs faster than real time.
It is measured with the whole video file available from the start, so that the model processes the video as fast as it can.
The processing time runs from the arrival of the video file to the moment when the last audio samples are available in CPU memory.
It covers video decoding, feature extraction, sampling and waveform decoding, but not model loading or writing the audio to a file.

\subsection{Subjective Evaluation}
\label{app:subjective}
The subjective evaluations of V2A and VT2A are each conducted with 20 participants on eight videos from the VGGSound test set.
Each video is shown with its ground-truth audio and with the audio generated by every compared model (Table~\ref{tab:subjective} for V2A and Table~\ref{tab:vt2a_subjective} for VT2A), in a random order for each participant and without the names of the models.
The rating interface for V2A is shown in Fig.~\ref{fig:listening}.
Every audio clip keeps the sample rate of its model, is cut to its first 8\,s and placed on the same 8\,s video track, and is normalized to a loudness of $-23$\,LUFS.
The participants are advised to use headphones in a quiet environment, and rate every clip on a five-point scale from 1 (bad) to 5 (excellent) in the aspects below.
For each model, we report the mean opinion score (MOS), the mean of its ratings, with its 95\% confidence interval.

\paragraph{MOS-Q.}
MOS-Q rates the quality and naturalness of the sound itself, judged by its clarity and by the presence of noise, distortion and artifacts.
The participants rate it without regard to the video.

\paragraph{MOS-T.}
MOS-T rates how well the sound is aligned in time with the actions in the video, judged by whether the sounds occur at the moments of events such as impacts, footsteps, the start of speech and the playing of instruments.

\paragraph{MOS-A.}
MOS-A rates how well the sound matches the text prompt, judged by whether the sound events described in the prompt are present in the audio.
It is rated only for VT2A, where the text prompt is shown with each video.

\section{Implementation Details}
\label{app:implementation}
\subsection{Model Architecture}
\label{app:architecture}
\begin{figure}[t]
    \centering
    \includegraphics[width=\linewidth]{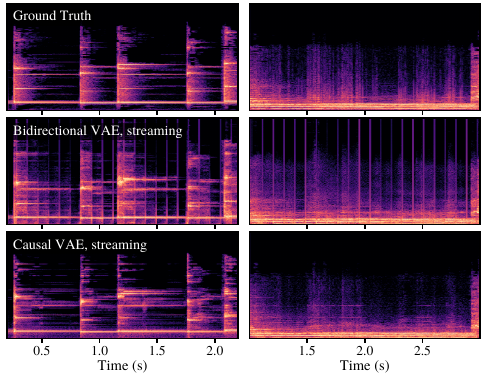}
    \caption{Ground truth and its streaming decoding with the bidirectional VAE of Stable Audio Open and with the causal VAE, where each block of about 0.1\,s is decoded without later latents.}
    \label{fig:vae_streaming}
\end{figure}

\paragraph{VAE.}
We use the causal variational autoencoder (VAE) of SoundReactor~\cite{saito2025soundreactor}, which follows the Oobleck architecture of Stable Audio Open~\cite{evans2025stable} and is kept frozen in all stages.
Its encoder downsamples 48\,kHz stereo audio with strides of 2, 4, 5, 5 and 8, a factor of 1600 in total, into 64-dimensional latents at 30\,Hz, and its decoder mirrors the encoder with transposed convolutions.
All convolutions are causal with padding only on the left, so each latent depends only on the current and past audio, and each audio sample only on the current and past latents.
SoundReactor trains the VAE with a multi-resolution STFT loss, adversarial and feature-matching losses, and a KL penalty on the latents.
During training, each latent is sampled from the VAE posterior, whose mean and variance are precomputed for every clip.
For streaming, the decoder is converted into the streaming decoder released with the VAE, whose convolutions keep their past inputs between calls, so each block of three latents is decoded into 0.1\,s of audio once it is generated.
A bidirectional VAE, in contrast, decodes a latent faithfully only together with the latents that follow it.
Fig.~\ref{fig:vae_streaming} decodes ground-truth latents block by block with the bidirectional VAE of Stable Audio Open, using all earlier latents but no later ones, which produces clicks at every block boundary.
With the causal VAE, streaming decoding gives the same waveform as decoding the whole clip.

\paragraph{DiT.}
The diffusion transformer (DiT) is built on the architecture of Stable Audio Open~\cite{evans2025stable}, with 24 layers, 24 attention heads, a width of 1536 and 1.17B parameters excluding the text encoder and the VAE.
Each VAE latent is projected to one token, so a 10\,s clip gives 300 tokens, which form 100 blocks of three tokens.
Each layer consists of self-attention with rotary position embeddings, cross-attention to the condition tokens and a feed-forward network.
The bidirectional and block-causal models share this architecture and differ only in their attention masks and noise-level tokens.

\paragraph{Video conditions.}
The video features are extracted as in MMAudio~\cite{cheng2025mmaudio}.
CLIP~\cite{radford2021learning} encodes frames sampled at 8\,fps and resized to $384\times384$ into 1024-dimensional features, which receive absolute position embeddings and enter the cross-attention together with the text.
A CLIP frame becomes available 0.04\,s after its timestamp, at the end of its display interval in the 25\,fps video, and an audio token attends to it only if it is available by the end of the slice that contains the start of the token.
Synchformer~\cite{iashin2024synchformer} runs on frames sampled at 25\,fps, resized to 224 on the short side and center-cropped.
Following LiveFoley~\cite{dai2026livefoley}, Synchformer encodes the trailing window that ends with each slice into a 768-dimensional feature once the slice has been played, which gives 125 features for a 10\,s clip.
The feature of the slice that contains the start of an audio token is projected into the modulation parameters of the adaptive layer normalization of that token.

\paragraph{Text and duration conditions.}
A frozen Flan-T5 base~\cite{chung2024scaling} encodes the text into at most 128 tokens.
As in Stable Audio Open, embeddings of the start time and the total duration enter the cross-attention and are added to every noise-level token.
A condition that a task does not provide is left empty, and the Synchformer features are set to zero for T2A.

\begin{figure*}[t]
    \centering
    \includegraphics[width=\linewidth]{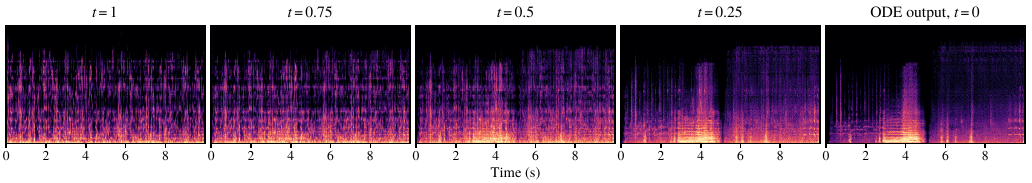}
    \caption{ODE trajectory of a training clip extracted with the Stage 2 model. Each spectrogram decodes the latents stored at one step time with the VAE, and the last one is the final output, which the Stage 3 model learns to predict from each of the other states.}
    \label{fig:ode_trajectory}
\end{figure*}

\subsection{Training Details}
\label{app:training}
\begin{table}[t]
    \centering
    \small
    \caption{Training hyperparameters of the four stages. LR denotes the learning rate, FP16 denotes mixed precision, and EMA denotes the exponential moving average of the weights.}
    \setlength{\tabcolsep}{3pt}
    \begin{tabularx}{\linewidth}{l *{4}{>{\centering\arraybackslash}X}}
        \toprule
        & Stage 1 & Stage 2 & Stage 3 & Stage 4 \\
        \midrule
        Iterations & 500k & 100k & 30k & 3k \\
        LR & $10^{-4}$ & $2{\times}10^{-6}$ & $2{\times}10^{-6}$ & $2{\times}10^{-6}$ \\
        LR of fake score & - & - & - & $4{\times}10^{-7}$ \\
        LR schedule & Inv.\ sqrt & Constant & Constant & Constant \\
        Batch size & 256 & 64 & 64 & 64 \\
        A100 GPUs & 32 & 8 & 8 & 8 \\
        Precision & FP16 & FP16 & FP16 & FP32 \\
        EMA decay & 0.9999 & - & - & - \\
        Condition dropout & 0.1 & 0.1 & - & - \\
        \midrule
        Clip length & \multicolumn{4}{c}{10\,s} \\
        Optimizer & \multicolumn{4}{c}{AdamW~\cite{loshchilov2017decoupled}} \\
        AdamW $\beta$ & \multicolumn{4}{c}{(0.9, 0.999)} \\
        Weight decay & \multicolumn{4}{c}{0.001} \\
        Gradient clipping & \multicolumn{4}{c}{None} \\
        \bottomrule
    \end{tabularx}
    \label{tab:training}
\end{table}

\paragraph{Common settings.}
We train all stages with the data and the task mixture described in Appendix~\ref{app:datasets}, and list their hyperparameters in Table~\ref{tab:training}.
Stage 2 is initialized from the exponential moving average (EMA) of the Stage 1 weights, and each later stage from the weights of the previous one.
The VAE and the condition encoders stay frozen, and only the DiT and the layers that embed the conditions are trained.

\paragraph{Stage 1.}
The noise level is drawn from the logit-normal distribution~\cite{esser2024scaling}, $t=\mathrm{sigmoid}(n)$ with $n\sim\mathcal{N}(0,1)$.
For classifier-free guidance (CFG)~\cite{ho2022classifier}, the video and text conditions of a sample are dropped together with probability 0.1.
The learning rate follows the schedule of Stable Audio Open~\cite{evans2025stable}, an exponential warmup followed by an inverse square-root decay, $10^{-4}\,(1-0.99^{s+1})\,(1+s/10^{6})^{-1/2}$ at step $s$.
The EMA of the weights with a decay of 0.9999 serves as the bidirectional DiT in all later stages.

\paragraph{Stage 2.}
We implement teacher forcing by packing a clean copy and a noisy copy of each clip into one sequence, each with one noise-level token per block.
Each noisy block attends to its own tokens and to the clean copies of the earlier blocks, so all blocks are trained in parallel with ground-truth history.
Every block draws its own logit-normal noise level, and the condition dropout of Stage 1 is kept.

\paragraph{Stage 3.}
The trajectories are extracted by solving Eq.~\ref{eq:sampling_ode} with the Stage 2 model, using 48 Euler steps on a uniform grid from $t=1$ to $t=0$, a guidance scale of 4.5 and full precision.
All blocks of a clip are denoised together with ground-truth history, and the 48 steps place the step times 1, 0.75, 0.5 and 0.25 of the four-step sampler on the grid.
We store the states at these four times and the final output for 131{,}072 training clips drawn with the task mixture of the training data.
Fig.~\ref{fig:ode_trajectory} shows the stored states of one clip.
During training, each sample draws one of the four step times for all its blocks, and no condition is dropped, since the guidance is contained in the trajectories.

\paragraph{Stage 4.}
The real score is given by the EMA weights of Stage 1 with a guidance scale of 4.5, and the fake score model is initialized from the same weights.
Each iteration updates the fake score model, and the generator is updated every five iterations, which gives 600 generator updates in 3k iterations.
Following Self Forcing~\cite{huang2026self}, the generator produces each clip block by block from its own history with the four-step sampler, the exit step is drawn uniformly for each clip and shared by all blocks, and only the output of this step receives gradients.
As in Causal Forcing~\cite{zhu2026causal}, the scores are evaluated at $t=5u/(1+4u)$ with $u\sim\mathcal{U}[0,1)$ and clamped to $[0.02,0.98]$, and the DMD gradient of Eq.~\ref{eq:dmd_grad} is normalized by the mean absolute difference between the generated clip and the prediction of the real score~\cite{yin2024one}.
The discriminator head follows Presto~\cite{novack2025presto}, in which four convolutional blocks with group normalization and SiLU, followed by a $1\times1$ convolution, map the features after the 12th block of the fake score model to one score for every four tokens.
Real and generated latents are noised at logit-normal noise levels before the head, which is trained with a least-squares loss.
The adversarial loss is weighted by 0.01 in the generator loss and by 0.005 in the fake score loss, and the head is updated with the fake score model at a learning rate of $10^{-4}$.

\subsection{Inference Details}
\label{app:inference}

\paragraph{Multi-step sampling.}
The bidirectional and block-causal DiTs are sampled with the Euler method on a uniform grid of 50 steps from $t=1$ to $t=0$.
For CFG, the unconditional prediction removes the conditions in the same way as the condition dropout in training, and the guidance scale is 4.5.
The bidirectional DiT uses the EMA weights and denoises the whole clip at once.
The block-causal DiT generates the blocks in order with 50 steps each, and every finished block is written into the KV caches of both predictions.

\paragraph{Few-step sampling.}
The models of Stages 3 and 4 use the four-step sampler with the step times 1, 0.75, 0.5 and 0.25 and no guidance.
The KV cache keeps all earlier blocks of a clip, so each block attends to the whole history.

\paragraph{Duration condition.}
Since the length of a stream is not known in advance, LiveAudio generates 10\,s of audio by default, which is cut to the duration of the video.

\begin{table*}[t]
    \centering
    \small
    \vspace{-0.5em}
    \caption{Video-text-to-audio results on the VGGSound test set with captions from VGGSound-Omni. As in Table~\ref{tab:main}, TTFF is reported only for streaming models. Results marked with $\dagger$ are quoted from the original papers. The teacher is our bidirectional DiT.}
    \setlength{\tabcolsep}{4.88pt}
    \begin{tabular}{l ccccc cc ccc cc}
        \toprule
        \multirow{2}{*}{Method} & \multicolumn{5}{c}{Distribution Matching} & \multicolumn{2}{c}{Audio Quality} & \multicolumn{3}{c}{Modality Alignment} & \multicolumn{2}{c}{Efficiency} \\
        \cmidrule(lr){2-6}\cmidrule(lr){7-8}\cmidrule(lr){9-11}\cmidrule(lr){12-13}
        & FD$_{\mathrm{O}}$$\downarrow$ & FD$_{\mathrm{V}}$$\downarrow$ & FD$_{\mathrm{P}}$$\downarrow$ & KL$_{\mathrm{S}}$$\downarrow$ & KL$_{\mathrm{P}}$$\downarrow$ & IS$_{\mathrm{P}}$$\uparrow$ & PQ$\uparrow$ & DeSync$\downarrow$ & IB$\uparrow$ & CLAP$\uparrow$ & TTFF (s)$\downarrow$ & RTF$\downarrow$ \\
        \midrule
        \multicolumn{13}{l}{\textit{Offline}} \\
        MMAudio & 69.01 & 0.83 & 4.22 & 1.41 & 1.62 & 14.90 & 5.99 & 0.447 & 0.331 & 0.373 & - & 0.38 \\
        AudioX & 73.47 & 1.24 & 8.01 & 1.55 & 1.57 & 15.14 & 6.29 & 1.235 & 0.266 & 0.409 & - & 1.35 \\
        ThinkSound & 47.12 & 1.15 & 8.89 & 1.55 & 1.72 & 10.51 & 6.06 & 0.555 & 0.255 & 0.358 & - & 0.71 \\
        HunyuanVideo-Foley & 80.65 & 1.90 & 10.30 & 1.44 & 1.78 & 11.39 & 6.19 & 0.534 & 0.319 & 0.364 & - & 1.52 \\
        Omni2Sound & 142.61 & 0.58 & 5.01 & 1.19 & 1.40 & 15.84 & 6.12 & 0.521 & 0.334 & 0.385 & - & 1.06 \\
        \rowcolor{oursrow} LiveAudio (teacher) & 67.66 & 1.28 & 4.33 & 1.33 & 1.44 & 14.87 & 6.25 & 0.498 & 0.323 & 0.327 & - & 0.52 \\
        \midrule
        \multicolumn{13}{l}{\textit{Streaming}} \\
        WorldSonus & 89.96 & 1.92 & 8.91 & 1.45 & 1.49 & 10.19 & 6.29 & 0.715 & 0.263 & 0.330 & 0.554 & 0.62 \\
        LiveFoley$^\dagger$ & - & - & 7.34 & - & 1.61 & 11.21 & 6.38 & 0.490 & 0.290 & - & - & - \\
        \rowcolor{oursrow} LiveAudio (ours) & 69.63 & 1.19 & 5.78 & 1.38 & 1.51 & 15.20 & 6.41 & 0.421 & 0.320 & 0.324 & 0.145 & 0.81 \\
        \bottomrule
    \end{tabular}
    \label{tab:vt2a}
\end{table*}

\paragraph{Streaming pipeline.}
The quality of LiveAudio is evaluated with the same streaming pipeline as its latency.
The video is decoded once in display order, and frames are sampled at 8\,fps for CLIP and 25\,fps for Synchformer, whose features are computed incrementally as the frames arrive.
Each block runs as two CUDA graphs, one for the four sampling steps and one for the pass that writes the KV cache, under BF16 autocast with the weights cast to BF16 in advance.
As soon as a block is finished, the streaming VAE decoder turns its three latents into 0.1\,s of 48\,kHz audio.

\section{Additional Results}
\label{app:results}

\subsection{Video-Text-to-Audio Generation}
\label{app:vt2a}
\enlargethispage{-\baselineskip}%
Table~\ref{tab:vt2a} reports VT2A results on the VGGSound test set, using the captions of VGGSound-Omni as the text prompts (Appendix~\ref{app:evaluation_data}).
We keep the baselines of Table~\ref{tab:main} that accept a text prompt with the video, namely MMAudio~\cite{cheng2025mmaudio}, AudioX~\cite{tian2025audiox}, Omni2Sound~\cite{dai2026omni2sound}, WorldSonus~\cite{fang2026worldsonus} and LiveFoley~\cite{dai2026livefoley}, and also compare with ThinkSound~\cite{liu2025thinksound} and HunyuanVideo-Foley~\cite{shan2025hunyuanvideo}, two offline models that take both inputs.
V-AURA~\cite{viertola2025temporally} and SoundReactor~\cite{saito2025soundreactor} support only streaming V2A and do not take the optional text input.
For ThinkSound, the caption is given in place of the chain-of-thought description that a multimodal language model would otherwise generate.
We quote the results of LiveFoley from its paper, which trains a text-conditioned variant for this task, and run all other baselines with their official checkpoints and inference settings, as in Table~\ref{tab:main}.
As in V2A, LiveAudio achieves the lowest DeSync and the highest PQ among all models.
It outperforms WorldSonus on most metrics and LiveFoley on all reported metrics, and is comparable to the offline models in distribution matching, audio quality and modality alignment.
For efficiency, LiveAudio runs faster than real time with a much lower TTFF than WorldSonus.
Table~\ref{tab:vt2a_subjective} reports the subjective evaluation, where MOS-A rates the alignment between the audio and the text prompt (Appendix~\ref{app:subjective}).
LiveAudio obtains the highest MOS-T among all models, and its MOS-Q and MOS-A are the highest among the streaming models and close to those of the best offline models.
\begin{table}[t]
    \centering
    \vspace{-0.5em}
    \small
    \caption{Subjective evaluation of video-text-to-audio generation. We report mean opinion scores with 95\% confidence intervals.}
    \setlength{\tabcolsep}{4pt}
    \begin{tabularx}{\linewidth}{l *{3}{>{\centering\arraybackslash}X}}
        \toprule
        Method & MOS-Q$\uparrow$ & MOS-T$\uparrow$ & MOS-A$\uparrow$ \\
        \midrule
        Ground Truth & 4.38$_{\pm0.17}$ & 4.43$_{\pm0.12}$ & 4.39$_{\pm0.16}$ \\
        \midrule
        \multicolumn{4}{l}{\textit{Offline}} \\
        MMAudio & 3.66$_{\pm0.17}$ & 3.98$_{\pm0.17}$ & 3.61$_{\pm0.17}$ \\
        AudioX & 3.51$_{\pm0.18}$ & 3.02$_{\pm0.18}$ & 3.21$_{\pm0.20}$ \\
        ThinkSound & 3.38$_{\pm0.21}$ & 3.70$_{\pm0.15}$ & 3.36$_{\pm0.29}$ \\
        HunyuanVideo-Foley & 3.60$_{\pm0.16}$ & 3.92$_{\pm0.14}$ & 3.53$_{\pm0.18}$ \\
        Omni2Sound & 3.80$_{\pm0.14}$ & 4.10$_{\pm0.12}$ & 3.81$_{\pm0.18}$ \\
        \midrule
        \multicolumn{4}{l}{\textit{Streaming}} \\
        WorldSonus & 3.55$_{\pm0.16}$ & 3.73$_{\pm0.18}$ & 3.48$_{\pm0.19}$ \\
        \rowcolor{oursrow} LiveAudio (ours) & 3.74$_{\pm0.19}$ & 4.11$_{\pm0.17}$ & 3.62$_{\pm0.25}$ \\
        \bottomrule
    \end{tabularx}
    \vspace{-1em}
    \label{tab:vt2a_subjective}
\end{table}

\subsection{Text-to-Audio Generation}
\label{app:t2a}
\begin{table*}[t]
    \centering
    \small
    \caption{Text-to-audio results on VGGSound-Omni and the evaluation set of Clotho v2. WorldSonus$^\ddagger$ is adapted from its official code.}
    \setlength{\tabcolsep}{2pt}
    \begin{tabularx}{\textwidth}{l *{12}{>{\centering\arraybackslash}X} c}
        \toprule
        \multirow{2}{*}{Method} & \multicolumn{6}{c}{VGGSound-Omni} & \multicolumn{6}{c}{Clotho v2} & Efficiency \\
        \cmidrule(lr){2-7}\cmidrule(lr){8-13}\cmidrule(lr){14-14}
        & FD$_{\mathrm{V}}$$\downarrow$ & FD$_{\mathrm{P}}$$\downarrow$ & KL$_{\mathrm{P}}$$\downarrow$ & IS$_{\mathrm{P}}$$\uparrow$ & PQ$\uparrow$ & CLAP$\uparrow$ & FD$_{\mathrm{V}}$$\downarrow$ & FD$_{\mathrm{P}}$$\downarrow$ & KL$_{\mathrm{P}}$$\downarrow$ & IS$_{\mathrm{P}}$$\uparrow$ & PQ$\uparrow$ & CLAP$\uparrow$ & RTF$\downarrow$ \\
        \midrule
        \multicolumn{14}{l}{\textit{Offline}} \\
        AudioLDM & 3.55 & 24.89 & 2.32 & 7.02 & 5.93 & 0.389 & 3.64 & 19.48 & 2.33 & 7.48 & 5.82 & 0.336 & 1.94 \\
        Tango & 3.71 & 21.36 & 1.96 & 6.61 & 6.07 & 0.395 & 5.20 & 27.33 & 2.43 & 6.18 & 6.24 & 0.281 & 1.43 \\
        AudioLDM 2 & 2.78 & 26.91 & 1.95 & 11.17 & 6.33 & 0.289 & 3.75 & 28.76 & 2.42 & 8.50 & 6.25 & 0.298 & 2.38 \\
        Stable Audio Open & 3.72 & 27.76 & 2.20 & 11.78 & 6.60 & 0.362 & 3.20 & 29.90 & 2.52 & 9.49 & 6.47 & 0.379 & 2.93 \\
        TangoFlux & 2.09 & 16.54 & 1.74 & 10.47 & 6.12 & 0.346 & 2.51 & 21.76 & 2.54 & 10.94 & 6.01 & 0.327 & 0.38 \\
        MMAudio & 1.56 & 8.98 & 2.03 & 14.62 & 5.99 & 0.377 & 2.37 & 15.58 & 2.50 & 9.34 & 6.11 & 0.381 & 0.17 \\
        AudioX & 1.57 & 10.62 & 1.60 & 15.75 & 6.29 & 0.382 & 2.86 & 21.39 & 2.61 & 11.38 & 6.08 & 0.336 & 1.01 \\
        \midrule
        \multicolumn{14}{l}{\textit{Streaming}} \\
        AudioGen & 2.37 & 13.91 & 2.10 & 11.00 & 5.49 & 0.287 & 2.42 & 19.82 & 2.61 & 7.43 & 5.34 & 0.288 & 1.75 \\
        WorldSonus$^\ddagger$ & 2.47 & 8.38 & 1.51 & 10.65 & 6.39 & 0.339 & 3.38 & 19.71 & 2.61 & 8.17 & 6.13 & 0.271 & 1.86 \\
        \rowcolor{oursrow} LiveAudio (ours) & 1.73 & 8.87 & 1.69 & 15.41 & 6.65 & 0.328 & 4.19 & 18.90 & 2.52 & 9.39 & 6.46 & 0.299 & 0.42 \\
        \bottomrule
    \end{tabularx}
    \label{tab:t2a_omni}
\end{table*}
\begin{table*}[t]
    \centering
    \small
    \caption{Ablation of the number of sampling steps in video-to-audio generation on the VGGSound test set. Each row is a separate generator that is distilled in Stage 4 from the same Stage 3 model with one, two or four sampling steps, and the four-step generator is LiveAudio.}
    \setlength{\tabcolsep}{2.5pt}
    \begin{tabularx}{\textwidth}{l *{10}{>{\centering\arraybackslash}X} cc}
        \toprule
        \multirow{2}{*}{Steps} & \multicolumn{5}{c}{Distribution Matching} & \multicolumn{2}{c}{Audio Quality} & \multicolumn{3}{c}{Modality Alignment} & \multicolumn{2}{c}{Efficiency} \\
        \cmidrule(lr){2-6}\cmidrule(lr){7-8}\cmidrule(lr){9-11}\cmidrule(lr){12-13}
        & FD$_{\mathrm{O}}$$\downarrow$ & FD$_{\mathrm{V}}$$\downarrow$ & FD$_{\mathrm{P}}$$\downarrow$ & KL$_{\mathrm{S}}$$\downarrow$ & KL$_{\mathrm{P}}$$\downarrow$ & IS$_{\mathrm{P}}$$\uparrow$ & PQ$\uparrow$ & DeSync$\downarrow$ & IB$\uparrow$ & CLAP$\uparrow$ & TTFF (s)$\downarrow$ & RTF$\downarrow$ \\
        \midrule
        1 & 83.31 & 2.13 & 10.18 & 2.04 & 2.13 & 13.96 & 6.22 & 0.405 & 0.298 & 0.246 & 0.123 & 0.57 \\
        2 & 73.42 & 1.32 & 7.70 & 1.96 & 2.12 & 15.16 & 6.27 & 0.408 & 0.318 & 0.252 & 0.129 & 0.64 \\
        \rowcolor{oursrow} 4 (ours) & 70.97 & 1.30 & 7.32 & 1.99 & 2.12 & 15.87 & 6.41 & 0.395 & 0.323 & 0.252 & 0.143 & 0.79 \\
        \bottomrule
    \end{tabularx}
    \label{tab:steps}
\end{table*}
In addition to the AudioCaps test set in Table~\ref{tab:t2a}, Table~\ref{tab:t2a_omni} evaluates T2A on VGGSound-Omni~\cite{dai2026omni2sound} and on the evaluation set of Clotho v2~\cite{drossos2020clotho}, both of which are held out from training.
We compare with the same baselines, namely AudioLDM~\cite{liu2023audioldm}, Tango~\cite{ghosal2023text}, AudioLDM~2~\cite{liu2024audioldm}, Stable Audio Open~\cite{evans2025stable}, TangoFlux~\cite{hung2024tangoflux}, MMAudio, AudioX, AudioGen~\cite{kreuk2022audiogen} and WorldSonus.
On VGGSound-Omni, each model generates audio from the caption of a test clip without the video.
On Clotho, whose clips are 15 to 30\,s long with five human-written captions each, each model generates 10\,s of audio from the first caption, which is compared with the first 10\,s of the ground truth.
WorldSonus~\cite{fang2026worldsonus} does not report T2A results, so we adapt its official code to this task.
Its results, marked with $\ddagger$, are generated with the video-free branch of its sampler, which uses learned null video tokens and runs offline rather than as a stream, and its RTF is measured on this branch.
On both benchmarks, LiveAudio remains comparable to the offline models and compares favorably with the streaming models, which is consistent with the results on AudioCaps.
For efficiency, we also report the RTF of generating a 10\,s clip, and LiveAudio also achieves real-time generation in T2A.{\widowpenalty=10000\par}

\section{Additional Ablation Studies and Analysis}
\label{app:ablation}

\subsection{Number of Sampling Steps}
\label{app:steps}
Table~\ref{tab:steps} studies the number of sampling steps of the generator.
We repeat Stage 4 with one, two and four sampling steps, where each generator is distilled from the same Stage~3 model under identical settings.
With two steps, the generator remains close to the four-step one in distribution matching and modality alignment, with slightly lower audio quality.
With a single step, DeSync stays on par with that of two steps, whereas distribution matching, audio quality and IB degrade clearly, so a single step mainly sacrifices generation quality rather than synchronization.
TTFF and RTF decrease only moderately with fewer steps, because the video encoding takes the same time regardless of the number of steps (Fig.~\ref{fig:latency}).
Since four steps already run faster than real time, we use them by default.

\subsection{Video Conditioning}
\label{app:conditioning}
\begin{table*}[t]
    \centering
    \small
    \caption{Comparison of video conditioning in video-to-audio generation on the VGGSound test set after each of the four training stages. (a) Standard conditioning in all stages. (b) Sync-aligned conditioning in all stages.}
    \vspace{-0.75em}
    \setlength{\tabcolsep}{3pt}
    \begin{tabularx}{\textwidth}{l *{10}{>{\centering\arraybackslash}X}}
        \toprule
        \multirow{2}{*}{Stages} & \multicolumn{5}{c}{Distribution Matching} & \multicolumn{2}{c}{Audio Quality} & \multicolumn{3}{c}{Modality Alignment} \\
        \cmidrule(lr){2-6}\cmidrule(lr){7-8}\cmidrule(lr){9-11}
        & FD$_{\mathrm{O}}$$\downarrow$ & FD$_{\mathrm{V}}$$\downarrow$ & FD$_{\mathrm{P}}$$\downarrow$ & KL$_{\mathrm{S}}$$\downarrow$ & KL$_{\mathrm{P}}$$\downarrow$ & IS$_{\mathrm{P}}$$\uparrow$ & PQ$\uparrow$ & DeSync$\downarrow$ & IB$\uparrow$ & CLAP$\uparrow$ \\
        \midrule
        \multicolumn{11}{l}{\textit{(a) Standard}} \\
        Stage 1 (Bidirectional) & 66.50 & 1.30 & 4.73 & 1.77 & 1.94 & 15.10 & 6.23 & 0.450 & 0.324 & 0.264 \\
        Stage 2 (Block-causal) & 76.20 & 2.63 & 9.76 & 2.14 & 2.21 & 10.55 & 6.35 & 0.610 & 0.282 & 0.225 \\
        Stage 3 (ODE) & 70.09 & 2.18 & 9.02 & 2.07 & 2.22 & 10.83 & 6.35 & 0.616 & 0.278 & 0.227 \\
        Stage 4 (DMD) & 75.68 & 2.15 & 9.35 & 1.91 & 2.09 & 12.77 & 6.43 & 0.488 & 0.303 & 0.248 \\
        \midrule
        \multicolumn{11}{l}{\textit{(b) Sync-aligned}} \\
        Stage 1 (Bidirectional) & 68.17 & 1.29 & 5.11 & 1.86 & 1.99 & 15.81 & 6.26 & 0.484 & 0.324 & 0.260 \\
        Stage 2 (Block-causal) & 68.19 & 1.44 & 6.79 & 2.08 & 2.15 & 12.54 & 6.24 & 0.535 & 0.307 & 0.235 \\
        Stage 3 (ODE) & 68.17 & 1.18 & 6.22 & 2.04 & 2.19 & 12.91 & 6.21 & 0.544 & 0.301 & 0.236 \\
        \rowcolor{oursrow} Stage 4 (DMD) & 70.97 & 1.30 & 7.32 & 1.99 & 2.12 & 15.87 & 6.41 & 0.395 & 0.323 & 0.252 \\
        \bottomrule
    \end{tabularx}
    \label{tab:conditioning}
\end{table*}
Table~\ref{tab:conditioning} reports the numerical results for Fig.~\ref{fig:conditioning}, where (a) uses the standard conditioning and (b) uses sync-aligned conditioning in all four stages.
The standard conditioning is the video conditioning of offline V2A models, for which we follow Omni2Sound~\cite{dai2026omni2sound}, where the features of the whole video are extracted before generation and the frames that an audio token receives are not restricted to those already played.
Synchformer encodes the video in segments of 16 frames with a stride of 8 frames, so the feature that modulates an audio token covers a whole segment and contains up to about 0.64\,s of video after the token, and the CLIP features of all frames are visible to every token in the bidirectional stage.
Sync-aligned conditioning injects the features through the same two paths, but gives each token only the frames that have been played (Appendix~\ref{app:architecture}).
In the bidirectional stage, the two are comparable, and the standard conditioning obtains a slightly lower DeSync owing to its lookahead.
From Stage~2 on, this lookahead is no longer available in streaming generation, and sync-aligned conditioning is clearly better in FD$_{\mathrm{V}}$, FD$_{\mathrm{P}}$, IS$_{\mathrm{P}}$ and DeSync, which confirms that sync-aligned conditioning~\cite{dai2026livefoley} is effective for streaming generation.
This is consistent with the conclusion in Sec.~\ref{sec:ablation}.

\subsection{Asymmetric DMD}
\label{app:dmd}
\begin{table*}[t]
    \centering
    \small
    \caption{Ablation of asymmetric DMD in video-to-audio generation on the VGGSound test set. Each variant changes one component.}
    \vspace{-0.75em}
    \setlength{\tabcolsep}{3pt}
    \begin{tabularx}{\textwidth}{l *{10}{>{\centering\arraybackslash}X}}
        \toprule
        \multirow{2}{*}{\makebox[0pt][l]{Variant}\hphantom{Stage 1 (Bidirectional)}} & \multicolumn{5}{c}{Distribution Matching} & \multicolumn{2}{c}{Audio Quality} & \multicolumn{3}{c}{Modality Alignment} \\
        \cmidrule(lr){2-6}\cmidrule(lr){7-8}\cmidrule(lr){9-11}
        & FD$_{\mathrm{O}}$$\downarrow$ & FD$_{\mathrm{V}}$$\downarrow$ & FD$_{\mathrm{P}}$$\downarrow$ & KL$_{\mathrm{S}}$$\downarrow$ & KL$_{\mathrm{P}}$$\downarrow$ & IS$_{\mathrm{P}}$$\uparrow$ & PQ$\uparrow$ & DeSync$\downarrow$ & IB$\uparrow$ & CLAP$\uparrow$ \\
        \midrule
        \rowcolor{oursrow} LiveAudio (ours) & 70.97 & 1.30 & 7.32 & 1.99 & 2.12 & 15.87 & 6.41 & 0.395 & 0.323 & 0.252 \\
        \textit{w/o adversarial loss} & 85.25 & 2.15 & 9.23 & 2.06 & 2.14 & 14.81 & 6.36 & 0.403 & 0.308 & 0.246 \\
        \textit{Symmetric DMD} & 94.84 & 2.71 & 11.96 & 2.25 & 2.31 & 12.49 & 6.38 & 0.502 & 0.269 & 0.219 \\
        \textit{Ground-truth history} & 191.82 & 8.12 & 44.49 & 3.60 & 3.08 & 5.76 & 6.02 & 0.711 & 0.147 & 0.103 \\
        \bottomrule
    \end{tabularx}
    \vspace{-0.5em}
    \label{tab:dmd}
\end{table*}
Asymmetric DMD is the stage that recovers the synchronization of the causal generator (Sec.~\ref{sec:ablation}), so we further examine its design in Table~\ref{tab:dmd} with three variants, each of which changes one component and is distilled from the same Stage~3 model under identical settings.
The first removes the adversarial loss, which shows what this loss contributes beyond distribution matching.
The second, symmetric DMD, takes the real score from the block-causal DiT of Stage~2 instead of the bidirectional DiT, with the fake score model initialized from the same weights, which tests whether the supervision of a bidirectional model is necessary.
The third generates each block from the ground-truth audio of the earlier blocks during training, as in Stages~2 and~3, instead of from the outputs of the generator, which tests whether the generator has to be trained on its own history.
Without the adversarial loss, DeSync changes little, whereas FD$_{\mathrm{O}}$, FD$_{\mathrm{V}}$, FD$_{\mathrm{P}}$ and IS$_{\mathrm{P}}$ degrade clearly.
This is consistent with Presto~\cite{novack2025presto} and AudioX-Turbo~\cite{tian2026audiox}, which add an adversarial loss on real audio to DMD to improve the perceptual quality of the audio, and indicates that in the DMD of audio models the adversarial loss mainly affects audio quality.
With symmetric DMD, all metrics degrade, and DeSync improves only slightly over that of Stage~3 (Table~\ref{tab:conditioning}), since the block-causal DiT cannot observe later audio or video either, which confirms that the synchronization is recovered by the supervision of the bidirectional DiT.
With ground-truth history, all metrics are far worse than those of Stage~3, which shows that distribution matching is effective only when it is applied to the audio that the generator produces from its own history~\cite{huang2026self}.

\subsection{Long-form Audio Generation}
\label{app:longform}
\begin{table}[t]
    \centering
    \footnotesize
    \caption{Long-form (60\,s) text-to-audio generation on the AudioCaps test set. WorldSonus$^\ddagger$ is adapted from its official code.}
    \setlength{\tabcolsep}{2pt}
    \begin{tabularx}{\linewidth}{l *{5}{>{\centering\arraybackslash}X} c >{\centering\arraybackslash}X}
        \toprule
        Method & FD$_{\mathrm{V}}$$\downarrow$ & FD$_{\mathrm{P}}$$\downarrow$ & KL$_{\mathrm{P}}$$\downarrow$ & IS$_{\mathrm{P}}$$\uparrow$ & PQ$\uparrow$ & CLAP$\uparrow$ & RTF$\downarrow$ \\
        \midrule
        AudioGen & 2.81 & 19.68 & 1.73 & 8.76 & 5.37 & 0.421 & 1.79 \\
        AudioLDM & 5.50 & 34.39 & 1.96 & 7.26 & 5.81 & 0.399 & 0.67 \\
        AudioLDM 2 & 6.17 & 36.49 & 1.87 & 8.55 & 5.96 & 0.376 & 0.62 \\
        WorldSonus$^\ddagger$ & 4.97 & 27.84 & 2.10 & 7.87 & 5.79 & 0.377 & 1.83 \\
        \rowcolor{oursrow} LiveAudio (ours) & 3.75 & 22.94 & 1.95 & 10.14 & 5.91 & 0.404 & 0.42 \\
        \bottomrule
    \end{tabularx}
    \label{tab:longform}
\end{table}
\begin{figure*}[t]
    \centering
    \includegraphics[width=\linewidth]{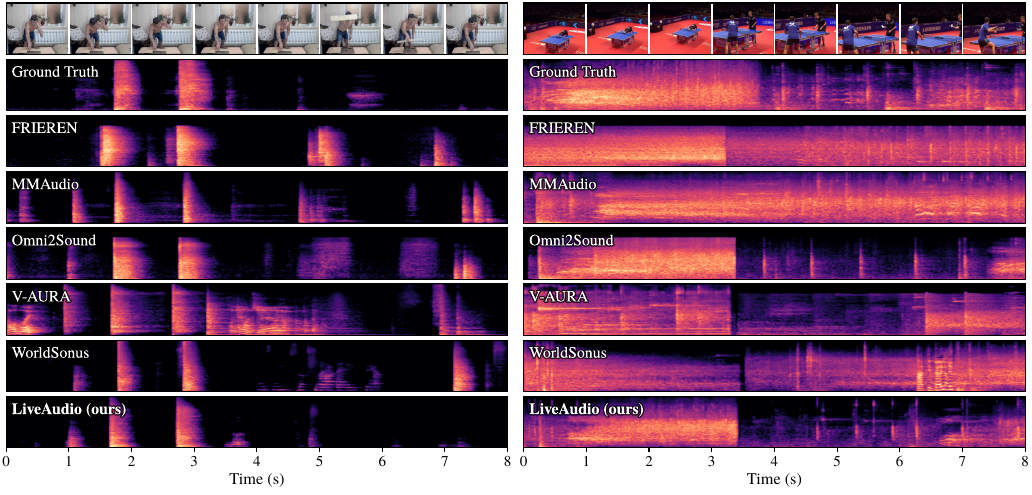}
    \caption{Qualitative comparison of video-to-audio generation on two test clips.}
    \label{fig:case_v2a}
    \belowtext
\end{figure*}
\begin{figure*}[t]
    \centering
    \includegraphics[width=\linewidth]{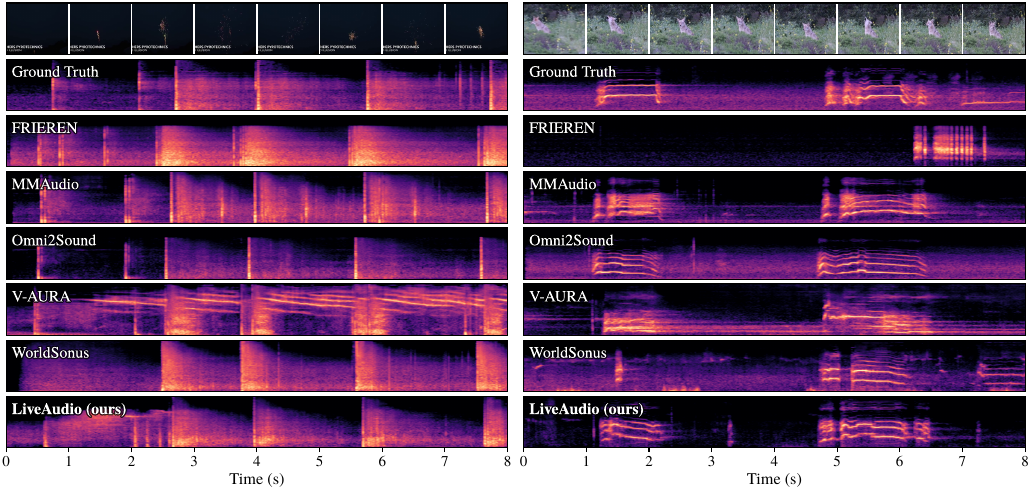}
    \caption{Qualitative comparison of video-to-audio generation on two more test clips.}
    \label{fig:case_v2a_more}
    \belowtext
    \vspace{30pt}
\end{figure*}
\begin{figure*}[t]
    \centering
    \includegraphics[width=\linewidth]{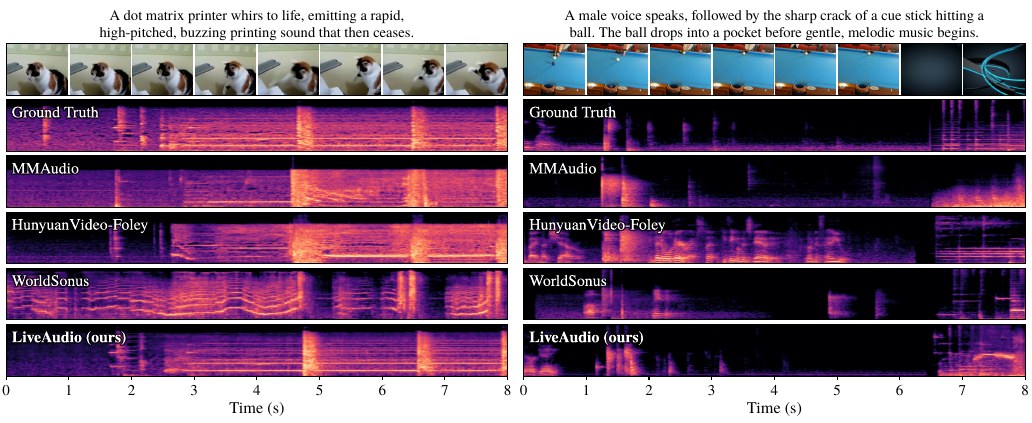}
    \caption{Qualitative comparison of video-text-to-audio generation on two test clips. The text prompt is shown above each clip.}
    \label{fig:case_vt2a}
\end{figure*}
\begin{figure*}[t]
    \centering
    \includegraphics[width=\linewidth]{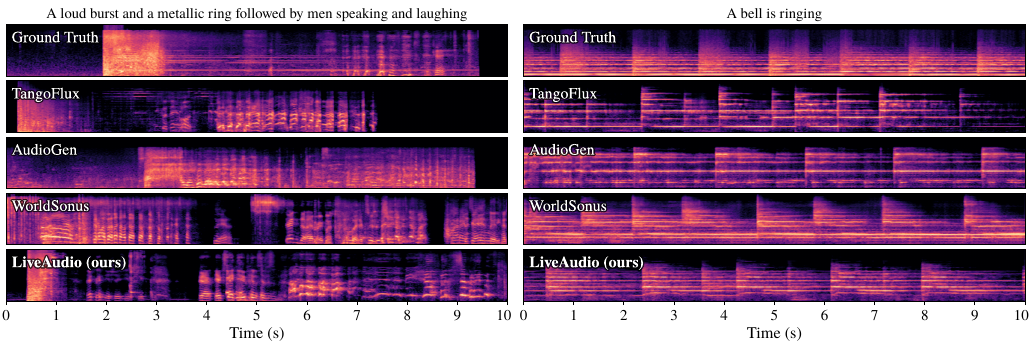}
    \caption{Qualitative comparison of text-to-audio generation on two captions of the AudioCaps test set. The text prompt is shown above each example.}
    \label{fig:case_t2a}
\end{figure*}
\begin{figure*}[t]
    \centering
    \includegraphics[width=\linewidth]{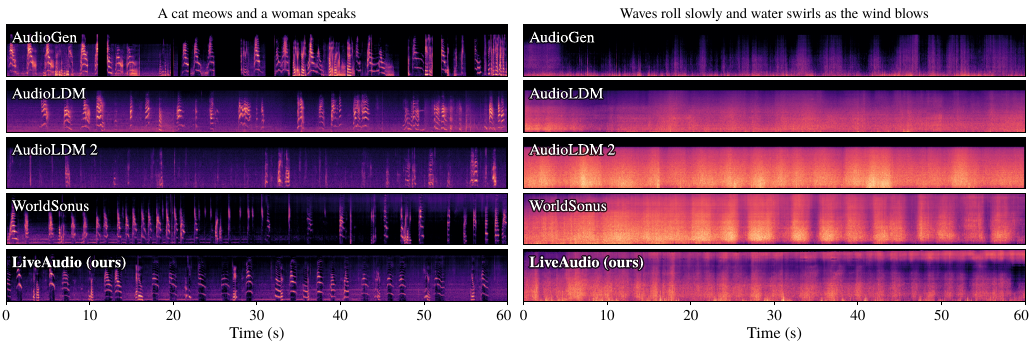}
    \caption{Qualitative comparison of long-form text-to-audio generation on two captions of the AudioCaps test set, where each model generates 60\,s of audio. The text prompt is shown above each example.}
    \label{fig:case_lft2a}
\end{figure*}
In streaming applications, the length of a stream is not known in advance and can far exceed the 10\,s clips on which LiveAudio is trained.
We therefore further explore the length extrapolation of LiveAudio in Table~\ref{tab:longform}, where it generates 60\,s of audio without additional training.
LiveAudio continues to generate block by block as a stream and keeps only the most recent 5\,s of audio in the KV cache, so that all blocks remain within the range seen in training.
AudioLDM~\cite{liu2023audioldm} extends the clip by outpainting, an offline procedure that denoises each 10\,s window as a whole, with its first half fixed to the audio already generated and its second half appended to the clip.
AudioLDM~2~\cite{liu2024audioldm} generates it at once, and AudioGen~\cite{kreuk2022audiogen} and WorldSonus~\cite{fang2026worldsonus} continue autoregressively with a sliding context.
We follow the common evaluation protocol for long-form audio generation~\cite{evans2024fast,evans2024long,jiang2025freeaudio,dai2025latent}.
LiveAudio obtains the highest IS$_{\mathrm{P}}$, is second only to AudioGen in FD$_{\mathrm{V}}$ and FD$_{\mathrm{P}}$, and outperforms WorldSonus on all metrics, with a KL$_{\mathrm{P}}$ close to those of the offline models.
This indicates that LiveAudio extends beyond its training length without additional training.

\section{Case Study}
\label{app:case_study}
We provide generated samples and comparisons with existing methods for V2A, VT2A and T2A on our demo page at \mbox{\url{https://streamaudio.github.io/}}.

Below, we provide additional spectrogram visualizations for V2A (Figs.~\ref{fig:case_v2a} and~\ref{fig:case_v2a_more}), VT2A (Fig.~\ref{fig:case_vt2a}), T2A (Fig.~\ref{fig:case_t2a}) and long-form T2A (Fig.~\ref{fig:case_lft2a}).

\end{document}